\documentclass[12pt]{article}

\usepackage{jheppub}
\usepackage{graphicx}
\usepackage{relsize}
\usepackage{tikz}
\usepackage{tikz-3dplot}
\usepackage{amsthm,amsmath,amssymb,bbm}
\usepackage{dsfont} 
\usepackage{enumitem} 
\usepackage{relsize}
\usepackage[utf8]{inputenc}
\usepackage{relsize}
\usepackage[T1]{fontenc} 
\usepackage{lmodern}
\usepackage{caption}
\usepackage{floatrow}
\usepackage{dcolumn}
\usepackage{bm}
\usepackage{color}
\usepackage[normalem]{ulem}
\usepackage{subfig}
\usepackage{mathtools}
\usepackage{scalerel,stackengine}
\usepackage{xcolor}
\usepackage[english]{babel}

\makeatletter

\define@key{x sphericalkeys}{radius}{\def\myradius{#1}}
\define@key{x sphericalkeys}{theta}{\def\mytheta{#1}}
\define@key{x sphericalkeys}{phi}{\def\myphi{#1}}
\tikzdeclarecoordinatesystem{x spherical}{%
    \setkeys{x sphericalkeys}{#1}%
    \pgfpointxyz{\myradius*cos(\mytheta)}{\myradius*sin(\mytheta)*cos(\myphi)}{\myradius*sin(\mytheta)*sin(\myphi)}}

\define@key{y sphericalkeys}{radius}{\def\myradius{#1}}
\define@key{y sphericalkeys}{theta}{\def\mytheta{#1}}
\define@key{y sphericalkeys}{phi}{\def\myphi{#1}}
\tikzdeclarecoordinatesystem{y spherical}{%
    \setkeys{y sphericalkeys}{#1}%
    \pgfpointxyz{\myradius*sin(\mytheta)*cos(\myphi)}{\myradius*cos(\mytheta)}{\myradius*sin(\mytheta)*sin(\myphi)}}

\define@key{z sphericalkeys}{radius}{\def\myradius{#1}}
\define@key{z sphericalkeys}{theta}{\def\mytheta{#1}}
\define@key{z sphericalkeys}{phi}{\def\myphi{#1}}
\tikzdeclarecoordinatesystem{z spherical}{%
    \setkeys{z sphericalkeys}{#1}%
    \pgfmathsetmacro{\Xtest}{sin(\tdplotmaintheta)*cos(\tdplotmainphi-90)*sin(\mytheta)*cos(\myphi)
    +sin(\tdplotmaintheta)*sin(\tdplotmainphi-90)*sin(\mytheta)*sin(\myphi)
    +cos(\tdplotmaintheta)*cos(\mytheta)}
    \pgfmathsetmacro{\ntest}{ifthenelse(\Xtest<0,0,1)}
    \ifnum\ntest=0
    \xdef\MCheatOpa{0.3}
    \else
    \xdef\MCheatOpa{1}
    \fi
    \pgfpointxyz{\myradius*sin(\mytheta)*cos(\myphi)}{\myradius*sin(\mytheta)*sin(\myphi)}{\myradius*cos(\mytheta)}}

\pgfdeclareplothandler{\pgfplothandlercurveto}{}{%
  point macro=\pgf@plot@curveto@handler@initial,
  jump macro=\pgf@plot@smooth@next@moveto,
  end macro=\pgf@plot@curveto@handler@finish
}

\def\pgf@plot@smooth@next@moveto{%
  \pgf@plot@curveto@handler@finish%
  \global\pgf@plot@startedfalse%
  \global\let\pgf@plotstreampoint\pgf@plot@curveto@handler@initial%
}

\def\pgf@plot@curveto@handler@initial#1{%
  \pgf@process{#1}%
  \pgf@xa=\pgf@x%
  \pgf@ya=\pgf@y%
  \pgf@plot@first@action{\pgfqpoint{\pgf@xa}{\pgf@ya}}%
  \xdef\pgf@plot@curveto@first{\noexpand\pgfqpoint{\the\pgf@xa}{\the\pgf@ya}}%
  \global\let\pgf@plot@curveto@first@support=\pgf@plot@curveto@first%
  \global\let\pgf@plotstreampoint=\pgf@plot@curveto@handler@second%
}

\def\pgf@plot@curveto@handler@second#1{%
  \pgf@process{#1}%
  \xdef\pgf@plot@curveto@second{\noexpand\pgfqpoint{\the\pgf@x}{\the\pgf@y}}%
  \global\let\pgf@plotstreampoint=\pgf@plot@curveto@handler@third%
  \global\pgf@plot@startedtrue%
}

\def\pgf@plot@curveto@handler@third#1{%
  \pgf@process{#1}%
  \xdef\pgf@plot@curveto@current{\noexpand\pgfqpoint{\the\pgf@x}{\the\pgf@y}}%
  \pgf@xa=\pgf@x%
  \pgf@ya=\pgf@y%
  \pgf@process{\pgf@plot@curveto@first}
  \advance\pgf@xa by-\pgf@x%
  \advance\pgf@ya by-\pgf@y%
  \pgf@xa=\pgf@plottension\pgf@xa%
  \pgf@ya=\pgf@plottension\pgf@ya%
  \pgf@process{\pgf@plot@curveto@second}%
  \pgf@xb=\pgf@x%
  \pgf@yb=\pgf@y%
  \pgf@xc=\pgf@x%
  \pgf@yc=\pgf@y%
  \advance\pgf@xb by-\pgf@xa%
  \advance\pgf@yb by-\pgf@ya%
  \advance\pgf@xc by\pgf@xa%
  \advance\pgf@yc by\pgf@ya%
  \@ifundefined{MCheatOpa}{}{%
  \pgf@plotstreamspecial{\pgfsetstrokeopacity{\MCheatOpa}}}
  \edef\pgf@marshal{\noexpand\pgfsetstrokeopacity{\noexpand\MCheatOpa}
  \noexpand\pgfpathcurveto{\noexpand\pgf@plot@curveto@first@support}%
    {\noexpand\pgfqpoint{\the\pgf@xb}{\the\pgf@yb}}{\noexpand\pgf@plot@curveto@second}
    \noexpand\pgfusepathqstroke
    \noexpand\pgfpathmoveto{\noexpand\pgf@plot@curveto@second}}%
  {\pgf@marshal}%
  \global\let\pgf@plot@curveto@first=\pgf@plot@curveto@second%
  \global\let\pgf@plot@curveto@second=\pgf@plot@curveto@current%
  \xdef\pgf@plot@curveto@first@support{\noexpand\pgfqpoint{\the\pgf@xc}{\the\pgf@yc}}%
}

\def\pgf@plot@curveto@handler@finish{%
  \ifpgf@plot@started%
    \pgfpathcurveto{\pgf@plot@curveto@first@support}{\pgf@plot@curveto@second}{\pgf@plot@curveto@second}%
  \fi%
}

\def\@fpheader{\relax}

\makeatother

\newcommand{\be}{\begin{equation}}
\newcommand{\ee}{\end{equation}}
\newcommand{\bea}{\begin{eqnarray}}
\newcommand{\eea}{\end{eqnarray}}

\newcommand{\tikzline}[2][fill=black]{\tikz[baseline=-0.5ex] \draw (0,0) circle ; \draw[line width=0.5mm, black] (0,0) -- (0.35,0); \draw (0.35,0) circle ;}%

\newcommand{\ket}[1]{| #1 \rangle}
\newcommand{\bra}[1]{\langle #1 |}
\newcommand{\scalar}[2]{\left\langle #1 | #2 \right\rangle}

\DeclareMathOperator{\Tr}{Tr}
\DeclareMathOperator{\Cof}{Cof}
\DeclareMathOperator{\arcsinh}{arcsinh}

\title{Effective geometry of Bell-network states on a dipole graph}

\author[a]{Bekir Bayta\c{s},}
\emailAdd{bekirbyts@gmail.com}
\author[b]{Nelson Yokomizo}
\emailAdd{yokomizo@fisica.ufmg.br}

\affiliation[a]{Department of Physics, {\.{I}}zmir Institute of Technology, G{\"{u}}lbah{\c{c}}e, Urla, 35430, {\.{I}}zmir, Turkey}
\affiliation[b]{Departamento de F\'isica - ICEx, Universidade Federal de Minas Gerais, CP 702, 30161-970, Belo Horizonte - MG, Brazil}

\abstract{Bell-network states are a class of entangled states of the geometry that satisfy an area-law for the entanglement entropy in a limit of large spins and are automorphism-invariant, for arbitrary graphs. We present a comprehensive analysis of the effective geometry of Bell-network states on a dipole graph. Our main goal is to provide a detailed characterization of the quantum geometry of a class of diffeomorphism-invariant, area-law states representing homogeneous and isotropic configurations in loop quantum gravity, which may be explored as boundary states for the dynamics of the theory. We found that the average geometry at each node in the dipole graph does not match that of a flat tetrahedron. Instead, the expected values of the geometric observables satisfy relations that are characteristic of spherical tetrahedra. The mean geometry is accompanied by fluctuations with considerable relative dispersion for the dihedral angle, and perfectly correlated for the two nodes.}

\begin{document}

\maketitle
\flushbottom

\newpage

\section{Introduction}

Loop quantum gravity (LQG) is a non-perturbative and background independent theory of quantum gravity~\cite{ashtekarjerry,rovellibook,thiemannbook,lqg30,lqgreviewebaa}. The Hilbert space of the theory, describing the quantum states of the geometry, can be described in a purely combinatorial manner, without reference to a manifold on which excitations of the gravitational field should manifest  \cite{rovellicombi}. In such a combinatorial approach, which underlies the current formulation of the spinfoam dynamics in LQG, the kinematical Hilbert space $\mathcal{K}$ of diffeomorphism-invariant states of the gravitational field is defined as a direct sum of Hilbert spaces $\mathcal{H}_{\Gamma}$ associated with abstract graphs $\Gamma$:
\be
\mathcal{H} = \bigoplus_\Gamma \, \mathcal{H}_{\Gamma}\,,
\ee
factored by an equivalence relation, $\mathcal{K} = \mathcal{H}/\!\!\sim$, that takes into account basic graph morphisms. Concretely, if a graph can be embedded in another graph,  $\Gamma' \subset \Gamma''$, then its Hilbert space is isomorphic to a subspace of that of the larger graph, $\mathcal{H}_{\Gamma'} \subset \mathcal{H}_{\Gamma''}$. Moreover, if two graphs are related by a graph isomorphism, then they have the same Hilbert space, i.e., isomorphic graphs are physically indistinguishable. In applications, one can truncate the theory to a single graph that captures the relevant degrees of freedom \cite{twisted-rovelli}. In cosmological applications, for instance, one restricts to a dipole graph to describe global properties of the geometry \cite{spinfoam-cosmo-flat,spinfoam-cosmo-lambda}.

If the theory is truncated to a single graph $\Gamma$, then the physical states $\ket{\Psi_\Gamma}$ and observables $\mathcal{O}_\Gamma$ defined on $\mathcal{H}_\Gamma$ must be invariant under the action of the automorphisms of the graph~\cite{BY-23}: 
\be
U_{\mathrm{A}} \ket{\Psi_\Gamma} = \ket{\Psi_\Gamma} \quad \,\, \textrm{and} \quad \,\, U_{\mathrm{A}} \, \mathcal{O}_\Gamma \, U^{-1}_{\mathrm{A}} = \mathcal{O}_\Gamma, \quad \, \forall \, \mathrm{A} \in \mathrm{Aut}(\Gamma)\,.
\ee
This condition ensures that the description of the states and observables depends only on the combinatorial structure of the graph, and not on the choice of a particular presentation, consisting of a discrete analogue of diffeomorphism invariance~\cite{ashtekarjerry,BY-23,arrighi, oriti-cola}. In highly symmetric graphs, automorphism invariance can impose severe restrictions on the truncated theory. In \cite{BY-23}, the special case of $2$-CH graphs was thoroughly discussed. Such graphs display the discrete analogue of the properties of homogeneity and isotropy, and provide a general framework for cosmological applications. General techniques for manipulating invariant objects on $2$-CH graphs were introduced in \cite{BY-23}, and applied, in particular, to the invariant definition of the entanglement entropy of a single node. 

On $2$-CH graphs, automorphism-invariant states describe homogeneous and isotropic geometries. Among these, one may expect to find special states that, in addition, are also semiclassical. These would be natural candidates to be employed in the formulation of a classical limit describing cosmological spacetimes from the general formalism of LQG. Among the properties naturally required from a viable semiclassical state in quantum gravity, beyond the condition that geometric observables are peaked on a classical configuration with small dispersions, is the presence of highly correlated fluctuations of the geometry, matching the picture for the quantum fluctuations of the geometry presented by semiclassical gravity as described by quantum field theory on curved spaces, expected to hold at scales larger than the Planck scale. Such correlations lead, in particular, to an area law for the entanglement entropy, and such a property was stressed in \cite{bianchi-myers} as a necessary condition of general validity for the emergence of a well-behaved semiclassical limit in any theory of quantum gravity.

Bell-network states were introduced in \cite{bell} as a family of highly entangled states of the geometry in LQG that solved certain gluing conditions required for the continuity of the metric, which essentially enforced that normals to glued faces are always aligned. The family of BN states on a fixed graph $\Gamma$ is parametrized by complex parameters $|\lambda_\ell| < 1$ assigned to the links of the graph. When they are chosen to be all equal, $\lambda_\ell=\lambda$, the BN state is expressed in the spin-network basis as:
\be
\ket{\Gamma,\mathcal{B}, \lambda} = \sum_{j_\ell, i_n} \, \prod_\ell \, q_{\lambda}(j_\ell) \,\, \overline{\mathcal{A}_\Gamma \left(j_\ell, i_n \right)} \,\, \ket{\Gamma, (j_\ell), (i_n)} \, ,
\label{eq:BN-state-intro}
\ee
where $q_{\lambda}(j_\ell)$ is the amplitude of finding the state $\ket{\Gamma,\mathcal{B},\lambda}$ with definite spins $j_\ell$, and $\mathcal{A}_\Gamma$ is the symbol of the graph $\Gamma$ \cite{bell}. For such a uniform choice of $\lambda_\ell$'s, it turns out that BN states are invariant under all automorphisms on any graph, as proved in~\cite{BY-23}. By projecting them into a subspace associated with definite spins $\{j_\ell\}$ in an invariant manner, one can also define invariant BN states with definite spins. In addition, BN states were shown to satisfy an area law for the entanglement entropy in the limit of uniformly large spins \cite{bia-dona-vilensky}, thus satisfying the criterion for semiclassicality proposed in \cite{bianchi-myers}. Therefore, the construction of the BN states (\ref{eq:BN-state-intro}), which was guided by a strategy of maximization of correlations for neighboring regions, ended up producing automorphism-invariant area-law states.

In this paper, we investigate in detail the effective geometry of BN states for a uniform choice of parameters $\lambda_\ell=\lambda$ on a simple $2$-CH graph, the dipole graph $\Gamma_{2,L}$, in an automorphism-invariant way. This setup provides an implementation of two main features expected from semiclassical states associated with cosmological spacetimes: (i) automorphism-invariance in a homogeneous and isotropic combinatorial context, and (ii) the presence of strong correlations in the fluctuations of the geometry. The application of the general techniques introduced in \cite{BY-23} to this concrete example provides an illustration of the operational implementation and physical meaning of such methods, and leads to an explicit characterization of the geometry of BN states on the dipole graph, required for their eventual application as boundary states for the LQG dynamics.

We begin by introducing group-averaging techniques for producing automorphism-invariant states and observables on $\Gamma_{2,L}$. These are then applied for the definition of the relevant set of invariant geometric observables---the volume, area, and angle operators---that describe the local geometry at a node. Next, we compute the expected values and dispersions of these operators for BN states with fixed spins, for varied choices of spins. Our analytical and numerical results indicate that the average geometry at each node, characterized by the expectation values of the one-node observables, is generally not that of a flat regular tetrahedron. Instead, the relations among the expected values of the geometric observables are those of a spherical regular tetrahedron. Only when the spins are equal ($j_\ell = j_0$), this geometry manifests as a flat tetrahedron. The dispersion of the mean geometry remains finite even for large spin values. Overall, our results reveal a non-trivial effective geometry displaying local curvature through the relations of the distinct geometric observables even at the level of a single node. The effective geometry corresponds to that of a $3$-sphere formed by the gluing of a pair of identical spherical regular tetrahedra, with perfectly correlated fluctuations.

The manuscript is organized as follows. We start by briefly reviewing the kinematical setting of loop quantum gravity in Section~\ref{sec:2}, which includes a discussion of automorphism-invariant states and observables on symmetric graphs. In Section~\ref{sec:3}, we present an overview of Bell-network states and introduce the special case of cosmological Bell-network (CBN) states that represent maximally entangled homogeneous and isotropic quantum geometries on $2$-CH graphs. The effective geometry of CBN states with definite spins on a dipole graph is determined in Section~\ref{sec:4}. In Section~\ref{sec:5}, we summarize the key findings of this work, and conclude by discussing potential avenues for further research.

\section{States and observables in loop quantum gravity}
\label{sec:2}

\subsection{Kinematical setting of loop quantum gravity}
\label{sec:2.1}

Let us briefly review the basic structure of loop quantum gravity \cite{ashtekarjerry,rovellibook,thiemannbook,lqg30,lqgreviewebaa}. The definition of the kinematical Hilbert space $\mathcal{K}$ of the quantized gravitational field in loop quantum gravity involves two steps. First, one considers the direct sum of Hilbert spaces of $SU(2)$-invariant states on generic abstract graphs $\Gamma$:
\[
\mathcal{H}=\bigoplus_\Gamma \mathcal{H}_\Gamma \, ,
\]
where the direct sum runs over all oriented graphs $\Gamma$. Second, one introduces an equivalence relation $\sim$ such that $\mathcal{K} = \mathcal{H}/\!\!\sim$, which we will discuss in the next section.

Let $L$ be the number of links and $N$ the number of nodes in a given graph $\Gamma$. The graph Hilbert space $\mathcal{H}_{\Gamma}$ associated with an oriented graph $\Gamma$,
\[
\mathcal{H}_{\Gamma} = L^2_{\Gamma}[\mathrm{SU(2)}^{L}/SU(2)^N] \, ,
\]
is the Hilbert space of gauge-invariant wavefunctions over a set of $SU(2)$ variables $h_\ell$ living at the links $\ell$ of the graph, defined by the invariance condition:
\begin{align}
\Psi_{\Gamma}(h_\ell) = \Psi_{\Gamma}(U_{s(\ell)} \, h_\ell \, U^{-1}_{t(\ell)})\,, \quad \forall U_{s(\ell)}, U_{t(\ell)} \in \mathrm{SU(2)} \, ,
\end{align}
where the labels $s(\ell)$ and $t(\ell)$ represent the source and target nodes of the link $\ell$. We denote the projector to the subspace of gauge-invariant function on the graph $\Gamma$ by:
\[
P_\Gamma: L^2_{\Gamma}[\mathrm{SU(2)}^{L}] \to \mathcal{H}_{\Gamma} \, .
\]

An orthonormal basis of $\mathcal{H}_{\Gamma}$ is provided by the family of spin-network states $\ket{\Gamma,(j_\ell), (i_n)}$, labeled by ordered sets of semi-integer spins $j_\ell=n/2$ at the links and $SU(2)$ intertwiners $i_n$ at the nodes of the graph. It is convenient to introduce the Hilbert space $\tilde{\mathcal{H}}_\Gamma \subset \mathcal{H}_\Gamma$ spanned by spin-network states with nonzero spins $j_\ell \neq 0$, $\forall \ell$.

The intertwiners $i_n$ form a basis of the intertwiner space of invariant tensors at the node $n$, defined as
\be
\mathcal{H}_n = \mathrm{Inv}_{\mathrm{SU(2)}} \left[ \left( \bigotimes_{s(\ell)= n} \, \mathcal{V}_{j_\ell} \right) \otimes \left( \bigotimes_{t(\ell)=n} \mathcal{V}_{j_\ell}\right) \right] \, ,
\label{eq:node-space}
\ee
where $\mathcal{V}_{j_\ell}$ is an irreducible representation of spin $j_\ell$ of $SU(2)$. 
In the tensor product above, representations of $SU(2)$ are associated with each link starting and ending at the node. The full state is obtained by taking the tensor product of the local intertwiners over all nodes of the graph:
\be
\ket{\Gamma, (j_\ell),(i_n)} = \bigotimes_n \ket{i_n} \,.
\ee

Geometrically, each node $n \in \Gamma$ represents a quantum polyhedron. The links at the node correspond to the faces of the polyhedron. Quantum polyhedra associated with distinct nodes of the graph are assembled together according to the graph structure: if two nodes are connected by a link, the corresponding polyhedra are joined along the faces described by the link. The graph can be seen as the dual graph to a triangulation, but a triangulation in which the individual polyhedra are now quantized.

The geometry of each polyhedron is described by the Penrose metric operator $g_{ab}(n)$. In order to define it, we first introduce, for each link $\ell$, the flux operator
\[
\vec {E}_\ell = \mathfrak{a}_{0} \vec J_\ell \, ,
\]
proportional to the left-invariant vector field $\vec J_\ell$ that acts as a derivative of the wavefunction $\Psi(h_1,\dots,h_L)$ with respect to the holonomy $h_\ell$. The constant $\mathfrak{a}_{0}=8\pi G\hbar \gamma$ is the area gap, written in terms of the Immirzi parameter $\gamma$. We are assuming that all links $\ell = na$ at the node $n$ point outwards from it\footnote{The choice of orientation of the links is arbitrary and carries no physical meaning. A given wavefunction defined with respect to some orientation is mapped into an equivalent wavefunction with respect to any other choice of orientation.}. Due to the $SU(2)$-invariance, intertwiners satisfy the closure relation:
\begin{equation}
\sum_{a=1}^{\mathrm{V}} \vec E_{na} \, \ket{i_n} =0 \, .
\end{equation} 
The Penrose metric operator at the node $n$ is defined as
\begin{equation}
\label{eq:shapeoperator}
g_{ab}(n) = \vec E_{na} \cdot \vec E_{nb} \, .
\end{equation}
The face areas and dihedral angles are expressed in terms of the metric operator as
\begin{align}
A_{na}&=\sqrt{g_{aa}(n)} \, , \nonumber \\
\cos \theta_{ab}(n) &= \frac{g_{ab}(n)}{\sqrt{g_{aa}(n)} \sqrt{g_{bb}(n)}} \, , \quad \text{for }a\neq b \,.
\label{eq:facearea-dihedral}
\end{align}

For a given configuration of the spins, all intertwiners $\ket{i_n} \in \mathcal{H}_n$ are eigenstates of the area operators,
\begin{equation}
A_{na} \, \ket{i_n} = \mathfrak{a}_{0}\,\sqrt{j_{na}(j_{na}+1)} \, \ket{i_n} \,,
\end{equation}
In addition, one can choose a linear basis of $\mathcal{H}_n$ that diagonalizes a specific operator at the node, as for instance the volume operator $V_n$. For an V-valent node, the Rovelli-Smolin volume operator \cite{rovelli-smolin,depietri} reads:
\be
V_n = \kappa_0 \,\, \frac{1}{2 \sqrt{2}} \,\, \sum_{1\leq a < b < c \leq \mathrm{V}} \sqrt{\left| \vec E_{na} \cdot (\vec E_{nb} \times \vec E_{nc}) \right| }\,,
\label{eq:volume-node}
\ee
where the constant $\kappa_0$ is a quantization ambiguity arising in the regularization of the classical formula for the volume density. This formula can be expressed in terms of the metric operator using the identity:
\[
\vec E_{na} \cdot (\vec E_{nb} \times \vec E_{nc}) = \frac{i}{\mathfrak{a}_0} \, [g_{ab}(n), g_{ac}(n)] \, .
\]

In the special case of a $4$-valent node, the classical formula for the volume of a tetrahedron in terms of the areas and normals of its faces can be used to fix this constant as $\kappa_0=2/3$, leading to: 
\be
V_n = \frac{\sqrt{2}}{3} \sqrt{\left| \vec E_{na} \cdot (\vec E_{nb} \times \vec E_{nc}) \right|} \, ,
\label{eq:volume-4-node}
\ee
where $na,nb,nc$ are any three links at the node $n$. The absence of a simple formula for the volume of general polyhedra with arbitrary number of faces prevents the same procedure to be directly implemented for nodes with arbitrary valence. An alternative definition of the volume operator that avoids such a quantization ambiguity was introduced in \cite{bianchi-polyhedra} on a coherent state representation.

\subsection{Invariant states and invariant observables on symmetric graphs}

A graph $\Gamma$ may have symmetries, as for instance those of a cubic lattice under translations and discrete rotations. In general, the notion of graph symmetry is captured by the concept of graph automorphism. In loop quantum gravity, physical states must be invariant under automorphisms, as a residual symmetry condition stemming from diffeomorphism invariance, when one considers states on a fixed graph. States and operators invariant under automorphisms in LQG have been thoroughly discussed in \cite{BY-23}. In this section, we present a brief review including the basic definitions and the relevant results for our purposes here.

An automorphism is a transformation of the graph that preserves its combinatorial structure. More concretely, let $\mathcal{N}(\Gamma)$ be the set of nodes of $\Gamma$. An automorphism of $\Gamma$ is a bijection $\mathrm{A}: \mathcal{N}(\Gamma) \to \mathcal{N}(\Gamma)$ that preserves adjacency relations. Employing the notation $r \sim s$ to specify that the nodes $r$ and $s$ are nearest neighbors connected by a link, an automorphism is characterized by the condition:
\be
r \sim s \iff \mathrm{A}(r) \sim \mathrm{A}(s) \, .
\ee
The automorphism group $\mathrm{Aut}(\Gamma)$ is the set of all automorphisms of the graph $\Gamma$. The action of an automorphism naturally extends from nodes to links: a link $\ell$ with source node $r$ and target node $s$ is simply mapped into a link $\mathrm{A}(\ell)$ with source node $\mathrm{A}(r)$ and target node $\mathrm{A}(s)$. 

The primary focus of graph theory is on the class of simple graphs, which are characterized by the absence of self-loops (a link connecting a node to itself) and multi-links (multiple links connecting the same pair of nodes); otherwise, the graph is referred to as a multigraph. Let $\mathcal{L}(\Gamma)$ be the set of links of $\Gamma$. For multigraphs, we define an autormorphism as a bijection $\mathrm{A}: (\mathcal{N}(\Gamma),\mathcal{L}(\Gamma)) \to (\mathcal{N}(\Gamma),\mathcal{L}(\Gamma))$ mapping nodes into nodes and links into links that preserves the graph. When convenient, we will represent the action of automorphisms on graphs and multigraphs simply as $\mathrm{A}: \Gamma \to \Gamma$, for brevity.

The action of an automorphism on a state $\ket{\Psi_\Gamma} \in \mathcal{H}_\Gamma$ is defined by:
\be
 \label{eq:action-automorphism}
\Psi_\Gamma(h_\ell) \mapsto (U_{\mathrm{A}} \Psi_\Gamma)(h_\ell) = \Psi_\Gamma(h_{\mathrm{A}(\ell)}) \, ,
\ee
i.e., the wavefunction is carried along as the links are shuffled around by the automorphism\footnote{When the orientation of the link $\mathrm{A}(\ell)$ disagrees with that of the link in $\Gamma$, we take $h_{\mathrm{A}(\ell)^{-1}}=h^{-1}_{\mathrm{A}(\ell)}$, where the link with the opposite orientation is represented by $\mathrm{A}(\ell)^{-1}$.}. In particular, the action of automorphisms on spin-network states reads \cite{BY-23}:
\be
\label{eq:autoaction}
U_{\mathrm{A}} \, \ket{\Gamma, (j_\ell), (i_n)} = (-1)^R \, \ket{ \Gamma, ( j'_\ell ), (i'_n)}  \, ,
\ee
where the configurations of spins and intertwiners are simply carried around by the automorphism, 
\be
j'_\ell = j_{\mathrm{A}^{-1}(\ell)} \, , \qquad i'_n = i_{\mathrm{A}^{-1}(n)} \, ,
\label{eq:carry-spins-intertwiners}
\ee
and $R$ is the number of links with semi-integer spins whose image under $\mathrm{A}$ has an orientation that disagrees with that of the graph. That is, the action of the transformation~\eqref{eq:action-automorphism} on a spin-network state produces a new spin-network state labeled by new spins and intertwiners, as described in Eqs.~\eqref{eq:autoaction} and \eqref{eq:carry-spins-intertwiners}.

We denote by $\mathcal{K}_{\Gamma} \subset \tilde{\mathcal{H}}_\Gamma \subset \mathcal{H}_\Gamma$ the subspace of all states with nonvanishing spins that are invariant under the automorphism group $\mathrm{Aut }(\Gamma)$, and the projector to this invariant subspace by $P_{A}: \mathcal{H}_{\Gamma} \to \mathcal{K}_{\Gamma}$. States in $\mathcal{K}_{\Gamma}$ will be called automorphism-invariant states, or simply invariant states. The full kinematical Hilbert space of LQG is then defined as
\be
\mathcal{K} = \bigoplus_{\Gamma} \mathcal{K}_\Gamma \, ,
\ee
where the sum runs over equivalence classes of isomorphic graphs $\Gamma$.

In the standard construction of the kinematical Hilbert space of loop quantum gravity through the quantization of the classical gravitational field in Ashtekar variables \cite{ashtekarjerry,rovellibook, thiemannbook}, automorphism invariance follows from the implementation of diffeomorphism invariance, and consists of a residual condition of diffeomorphism invariance, after the restriction to states based on a given graph. The space of solutions to the kinematical constraints of loop quantum gravity on a graph $\Gamma$, describing states invariant under gauge transformations and spatial diffeomorphisms, is the Hilbert space $\mathcal{K}_\Gamma$. Physical states must, in addition, satisfy the Hamiltonian constraint.

At the graph level, automorphism-invariance establishes the independence of the quantum geometry of a state based on a graph $\Gamma$ on the choice of a concrete presentation of the graph. More explicitly, consider an automorphism that acts on links as $\mathrm{A} (\ell_i)=\ell_j$. The automorphism induces a permutation of the labelling of the links, $\ell_i \mapsto \ell_j$. It turns out that, for an automorphism-invariant state $\Psi_\Gamma(h_\ell) \in \mathcal{K}_\Gamma$, such a relabelling is irrelevant, as the form of the wavefunction is preserved, $\Psi_\Gamma(h_\ell) = U_{\mathrm{A}} \Psi_\Gamma(h_\ell)=\Psi_\Gamma(h_{\mathrm{A}(\ell)})$. The distinct labellings can thus be seen as a gauge freedom, associated with the choice of an arbitrary concrete presentation of the graph, which does not affect the quantum geometry described by the state. In this work, we will mostly work with $SU(2)$-invariant states, and reserve the terms gauge-invariant, gauge-fixed, etc., to describe properties of states or observables under autormorphisms, for convenience. 

We say that an observable $\mathcal{O}: \mathcal{H}_{\Gamma} \to \mathcal{H}_{\Gamma}$ is gauge-invariant, or automorphism-invariant, when it remains unchanged under the action of arbitrary automorphisms,
\be
\mathcal{O} \mapsto U_{\mathrm{A}} \, \mathcal{O} \, U_{\mathrm{A}}^{-1}=\mathcal{O} \, , \quad \forall \, \mathrm{A} \, .
\ee
If $\mathcal{O}$ is a gauge-dependent operator defined on a specific presentation of a graph $\Gamma$, then we can introduce a gauge-invariant operator associated with $\mathcal{O}$ through an averaging procedure:
\be
\mathcal{O}_{\mathrm{inv}} = \frac{1}{|\mathrm{Aut }(\Gamma)|} \, \sum_{\mathrm{A}} U_{\mathrm{A}} \, \mathcal{O} \, U_{\mathrm{A}}^{-1} \, .
\ee
Similarly, for any gauge-dependent state $\ket{\Psi_\Gamma}$, an associated invariant state $\ket{\Psi_\Gamma}_{\mathrm{inv}}$ can be obtained by projecting to the automorphism-invariant subspace through an averaging procedure:
\bea
\ket{\Psi_\Gamma}_{\mathrm{inv}} &=& P_{\mathrm{A}} \, \ket{\Psi_\Gamma} \nonumber \\
	&=& \frac{1}{|\mathrm{Aut}(\Gamma)|} \, \sum_{\mathrm{A} \in \mathrm{Aut}(\Gamma)} \!\! U_{\mathrm{A}} \, \ket{\Psi_\Gamma} \, .
\label{eq:projection-averaging}
\eea

\subsection{Homogeneity and isotropy on graphs}

The conditions of homogeneity and isotropy in quantum geometry can be established by selecting an appropriate class of graphs and imposing restrictions on the state space and the algebra of observables to include only those elements that remain invariant under the automorphism group of the graph.

\begin{figure}
\centering
 \includegraphics[scale=0.65]{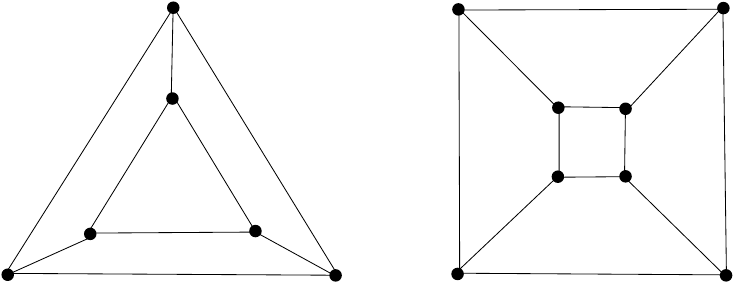}
\caption{Left: A node-transitive (1-CH) graph, the 3-prism. Right: An arc-transitive (2-CH) graph, the cubical graph.}
\label{fig:somegraph}
\end{figure}

The symmetries of an abstract graph are determined by its automorphism group, with each element of this group preserving the graph's structural properties, such as adjacency, valency, and distance. An asymmetric graph is characterized by an automorphism group that consists solely of the trivial automorphism, namely, the identity mapping of the graph. In contrast, certain classes of symmetric graphs are distinguished by the specific properties of their non-trivial automorphism groups~\cite{biggs}. In this work, we are interested in the class of so-called 2-CH graphs~\cite{BY-23}.

A simple class of graphs possessing non-trivial automorphisms is referred to as \textit{node-transitive} or \textit{1-CH}.  A graph $\Gamma_H$ is classified as 1-CH if, for any two nodes $n, n' \in N(\Gamma_H)$, there exists an automorphism $\mathrm{A}$ such that $\mathrm{A}(n) = m$; equivalently, if the automorphism group $\mathrm{Aut}(\Gamma_H)$ acts transitively on the nodes of $\Gamma_H$. So, in a 1-CH graph, all nodes are equivalent. This characteristic provides a notion of homogeneity on graphs. Two examples of 1-CH graphs are depicted in Fig.~\ref{fig:somegraph}.

\begin{figure}[hbtp]
\centering
 \includegraphics[scale=0.85]{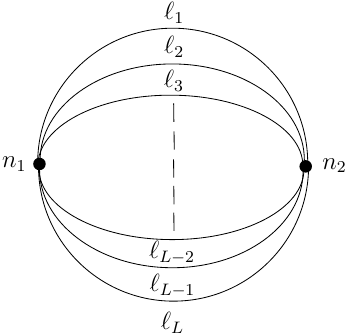}
\caption{The dipole graph $\Gamma_{2,L}$ with $L$ links.}
\label{fig:dipolegraph}
\end{figure}

1-CH graphs are a subset within the broader classification of $k$-connected homogeneous ($k$-CH) graphs. A graph is designated as $k$-CH if any isomorphism between connected induced subgraphs of order at most $k$ extends to an automorphism of the entire graph. In particular, in a \textit{2-CH} or \textit{arc-transitive} graph, for any two pairs of adjacent nodes $n_1 \sim n'_1$ and $n_2 \sim n'_2$, there exists an automorphism $\mathrm{A}$ such that
\be
\mathrm{A}(n_1) = n_2 \quad \mathrm{and} \quad \mathrm{A}(n'_1) = n'_2 \, . 
\ee
A 2-CH graph $\Gamma_C$ is inherently also a 1-CH graph, as are all $k$-CH graphs for $k \geq 1$. Henceforth, all links are equivalent in a 2-CH graph and this provides a notion of isotropy on graphs (see Fig.~\ref{fig:somegraph}). More generally, in the context of multigraphs, we define $\Gamma_C$ as 2-CH if, for any pair of oriented links $\ell,\ell'$, there exists an automorphism $\mathrm{A}$ such that $\mathrm{A}(\ell)=\ell'$. The simplest example of a 2-CH multigraph is the dipole graph $\Gamma_{2,L}$, depicted in Fig.~\ref{fig:dipolegraph}.

Homogeneity and isotropy in a quantum geometry can be associated with the states and the algebra of observables that remain invariant under the automorphism groups of 1-CH graphs  $\Gamma_H$ and 2-CH  $\Gamma_C$ graphs, respectively.

The nodes of a 1-CH graph $\Gamma_H$, which constitute the building blocks of quantum geometry, are equivalent.  Consequently, the quantum geometry associated with an automorphism-invariant state $\ket{\Psi_{\Gamma_H}} \in \mathcal{K}_{\Gamma_H}$ does not differentiate between nodes of the graph. We refer to such invariant states $\ket{\Psi_{\Gamma_H}}$ on $\Gamma_H$ as \textit{homogeneous} states.

Since any two links connected to a given node in a 2-CH graph $\Gamma_C$ can be related by an automorphism that preserves the node, the invariant states on $\Gamma_C$ exhibit a quantum analogue of isotropy: the distinct links at a node represent directions emanating from that node. Analogous to the space of homogeneous states $\ket{\Psi_{\Gamma_H}}$, we call the states $\ket{\Psi_{\Gamma_C}} \in \mathcal{K}_{\Gamma_C}$ on $\Gamma_C$ \textit{cosmological} states. A general class of LQG states exhibiting these symmetries, which serve as quantum analogues of homogeneity and isotropy, was introduced in~\cite{BY-23}.

\section{Entangled invariant states: Bell-network states}
\label{sec:3}

Bell-network (BN) states were originally introduced in \cite{bell} as a class of highly entangled states in loop quantum gravity. The main motivation for the analysis in \cite{bell} was an observed interplay between an information-theoretic property---maximal entanglement---and a special property of smoothness of the quantum geometry. It was later recognized that the entanglement between the shapes of quantum polyhedra in Bell-network states also gives rise to an area law for the entanglement entropy in the regime of uniformly large spins \cite{bia-dona-vilensky}. As argued in \cite{bianchi-myers}, an area law constitutes a necessary condition for the selection of semiclassical states in a theory of quantum gravity, ensuring that the state is able to reproduce the correlations of fluctuations of the geometry present in the continuum. Both such properties of Bell-network states are desirable from the point of view of a semiclassical limit, as described by a smooth geometry perturbed by strongly correlated vacuum quantum fluctuations.

A generic Bell-network state $\ket{\Gamma,\mathcal{B}, (\lambda_\ell)} $ is labeled by a set of complex parameters $\lambda_\ell$ living at the links of the graph $\Gamma$, with $|\lambda_\ell| <1$. When these parameters are all equal $\lambda_\ell = \lambda$, the states are invariant under the action of graph automorphisms on arbitrary graphs \cite{BY-23}. In particular, for the case of the highly symmetric $2$-CH graphs, discrete quantum analogues of the conditions of homogeneity and isotropy then hold \cite{BY-23}. BN states on $2$-CH graphs can then be seen as concrete examples of spatial sections of homogeneous and isotropic quantum spaces defined for a large class of graphs in loop quantum gravity.

The main goal of this work is to clarify the properties of the effective geometry described by BN states. For this purpose, we will compute the average and dispersion of the main geometric observables for BN states on a simple $2$-CH graph---the dipole graph---, taking into account automorphism invariance. In this section, we briefly review the definition and basic properties of Bell-network states on $2$-CH graphs, and then discuss the projections of the states to subspaces specified by a given spin configuration, which will be of interest later. In the Section~\ref{sec:4}, we present the results of our analysis of the effective geometry of BN states on the dipole graph.

\subsection{Bell-network states on a generic graph}

Bell-network states were originally constructed through the application of squeezed vacua techniques to the bosonic representation of LQG \cite{Girelli,Borja,Livine:2011gp,squeezedvacua} (for a review on the bosonic representation, see \cite{bosonic}). Their explicit form can also be described in terms of the spin states $\ket{j,m}$, as follows. For each link $\ell$, we first define the local state:
\be
\ket{\mathcal{B}, \lambda_\ell}_\ell= (1-|\lambda_\ell|^2) \, \sum_{j_\ell} \lambda_\ell^{2j_\ell} \, \sqrt{2j_\ell+1} \, \ket{\mathcal{B},j_\ell} \, ,
\ee
where $|\lambda_\ell|<1$, and
\be
\ket{\mathcal{B},j} = \frac{1}{\sqrt{2j+1}} \sum_{m=-j}^j (-1)^{j-m} \ket{j,m}_s \ket{j,-m}_t
\ee
is the maximally entangled state of spin $j$. By taking the tensor product of such link-states over a graph $\Gamma$ and projecting to the $SU(2)$-invariant subspace, we obtain a state on the kinematical Hilbert space $\mathcal{H}_\Gamma$,
\be
\ket{\Gamma,\mathcal{B}, (\lambda_\ell)} = P_\Gamma \bigotimes_\ell  \ket{\mathcal{B}, \lambda_\ell}_\ell \, ,
\label{eq:bell-state}
\ee
which we call a Bell-network state on the graph $\Gamma$. The state depends on the set of parameters $\lambda_\ell$ associated with the links, which can be chosen independently over the graph. For the case $\lambda_\ell = \lambda$, the Bell-network state is invariant under the action of all automorphisms $A$ for arbitrary graphs $\Gamma$~\cite{BY-23}:
\be
U_A \, \ket{\Gamma,\mathcal{B}, \lambda} = \ket{\Gamma,\mathcal{B},\lambda}\,, \quad \forall \, A \in \mathrm{Aut}(\Gamma) \,.
\ee

The projection $P_\Gamma$ can be implemented using the resolution of the identity in the spin-network basis,
\be
P_\Gamma=\sum_{j_\ell, i_n} \ket{\Gamma, (j_\ell), (i_n)} \bra{\Gamma, (j_\ell), (i_n)} \, ,
\ee
leading to a formula for the Bell-network states as an expansion over spin configurations,
\begin{equation}
\label{BNdefG}
\ket{\Gamma,\mathcal{B},\lambda} = \sum_{j_\ell} \, \prod_\ell \, q_{j_\ell}(\lambda) \, \ket{\Gamma,\mathcal{B}, (j_\ell)} \, ,
\end{equation} 
with expansion coefficients
\be
q_{j_\ell}(\lambda)  = \left( 1- \left| \lambda \right|^2 \right) \lambda^{2 j_\ell}  \sqrt{2 j_\ell +1} \, .
\ee
Bell-network states at fixed spin configuration $(j_\ell)$ are given by:
\begin{equation}
\label{BNdef}
\ket{\Gamma, \mathcal{B}, (j_\ell)} = \frac{1}{\sqrt{\mathcal{N}}} \, \sum_{i_n} \, \overline{\mathcal{A}_\Gamma \left(j_\ell, i_n \right)} \,\, \ket{\Gamma, (j_\ell),(i_n)} \, .
\end{equation}
The quantity $\mathcal{N}$ is a normalization constant, and the amplitude $\mathcal{A}$ is the $SU(2)$-symbol of the graph $\Gamma$~\cite{bell}. This completes the definition of the states. Let us comment on some of their general features.

\paragraph*{Entanglement as a probe for semiclassical geometry.} The Bianchi-Myers conjecture~\cite{bianchi-myers} states that a defining feature of semiclassical states within any quantum gravity framework is their adherence to an area law for the entanglement entropy. In loop quantum gravity, the entanglement entropy associated with a region typically scales as the volume of the region \cite{bianchi-livine}, and area-law states constitute a very special family of states, to which one could restrict for the identification of semiclassical states, according to the conjecture. BN-states were explicitly shown through analytical and numerical results~\cite{bia-dona-vilensky} to satisfy an area law for the entanglement entropy in the limit of uniformly large spins.

\paragraph*{Gluing conditions.} The entanglement between the intertwiner degrees of freedom for BN-states enforces gluing conditions for the quantum geometry of neighboring polyhedra. States of the geometry in loop quantum gravity correspond to wavefunctions over a space of twisted geometries \cite{twisted-FS,twisted-rovelli}, a generalization of Regge geometries that allow for discontinuities of the metric at the boundaries of glued polyhedra. Such discontinuities are not present in classical geometries. In Bell-network states, the strong correlations of the fluctuations of the geometry tame the discontinuities of twisted geometries, restricting the fluctuations of the geometry to a subspace of better behaved vector geometries \cite{barrett-2009,barrett-2010}, in which the normals of neighboring polyhedra are glued back-to-back.

\paragraph*{Uniform construction on graphs with higher complexity.} A Bell-network state at fixed spin configuration~(\ref{BNdef}) represents a superposition of spin-network states. Each state is weighted by the $SU(2)$ symbol of the corresponding graph, which solely encodes the combinatorial structure of the graph along with the quantum degrees of freedom intrinsic to the theory, without necessitating any additional structure or data.  In contrast with coherent states~\cite{livine-coherent}, the construction of Bell-network states does not require the specification of a local classical discrete geometry at each node on which the state would be peaked on. 

\paragraph*{Applications of Bell-network states.} BN-states can be considered as Hartle-Hawking states for the early universe in spinfoam cosmological models~\cite{vidotto1}. The mean geometry, fluctuations, and correlations of these states were computed on finite graphs (duals of the boundary of a 4-simplex and a 16-cell~\cite{vidotto2, vidotto3}) using transition amplitudes. Bell-network states have also been used for the computation of the vertex amplitude of a pentagram using quantum algorithms~\cite{jakub}.

\subsection{Bell-network states on $\Gamma_C$ with fixed spins}

As observed in the previous section, Bell-network states $\ket{\Gamma,\mathcal{B}, (\lambda_\ell)}$ with a uniform set of parameters $\lambda_\ell = \lambda$ are automorphism-invariant. Therefore, on $2$-CH graphs, they are cosmological states. Accordingly, we refer to Bell-network states on $2$-CH graphs as cosmological Bell-network (CBN) states. We denote them by:
\be
\ket{\Gamma_C, \mathcal{B}, \lambda} :=  P_{\Gamma_C} \bigotimes_\ell  \,  \ket{\mathcal{B}, \lambda}_\ell \,.
\label{eq:cbn-def}
\ee
The average local geometry of the state $\ket{\Gamma_C, \mathcal{B}, \lambda}$ is the same at all nodes and does not distinguish between links at any node. The fluctuations of the quantum geometry include, nonetheless, fluctuations over inhomogeneous and anisotropic configurations. For instance, the average area at each link is always the same, but the state has nonzero projections on spin-network states with nonuniform configurations of spins $j_\ell$, implying that distinct areas can be measured at distinct links. The probabilities are such that the anisotropies are averaged out for the full state.

From the CBN-states defined in Eq.~(\ref{eq:cbn-def}), one can obtain new cosmological states by projecting to subspaces associated with definite spins $\{j_\ell\}$, and then applying the projector to the invariant space $\mathcal{K}_{\Gamma_C}$, as described in Eq.~(\ref{eq:projection-averaging}):
\be
\ket{\Gamma_C, \mathcal{B},\{j_\ell\}} := P_{\mathrm{A}} \, P_{(j_\ell)} \, \ket{\Gamma_{C},\mathcal{B},\lambda} \,. 
\label{eq:gamma-fixed-j}
\ee 
As automorphisms permute the spins at the links, the result of the projection will involve a superposition of states with distinct spin configurations $(\tilde{j_\ell}=j_{\mathrm{A}^{-1}(\ell)})$ on the graph, including all permutations of spins that can be reached by automorphisms from an initial configuration. We call such projections CBN-states with definite spins.

The simplest case of CBN states with definite spins corresponds to the case of a uniform configurations of spins, $j_\ell = j_0, \, \forall \, \ell$. In this case, the action of $P_{\mathrm{A}}$ is trivial. For the case of distinct spins, the CBN state with fixed spins $j_\ell$ involves a superposition of spin-networks with all possible configurations $(j_{\mathrm{A}^{-1}(\ell)})$ that can be reached by the action of automorphisms on the specified configuration $j_\ell$.

\section{Effective geometries of Bell-network states on $\Gamma_{2,L}$}
\label{sec:4}

In this section, we study the effective geometry of Bell-network states on a dipole graph $\Gamma_{2,L}$. We first describe the space of states and the invariant observables on $\Gamma_{2,L}$, and then discuss the properties of Bell-network states at fixed spins on this graph. Next, we explicitly determine average values and dispersions of basic geometric observables for Bell-network states, focusing on configurations with fixed spins and four links on $\Gamma_{2,4}$.

\subsection{Invariant states and invariant observables on $\Gamma_{2,L}$}
\label{subsec:dipole-graph}

The dipole graph $\Gamma_{2,L}$ is a homogeneous multi-graph without self-loops consisting of two nodes connected by $L$ links (see Fig.~\ref{fig:dipolegraph}).

\paragraph*{Invariant states on $\Gamma_{2,L}$.} 

An orthonormal basis for the Hilbert space $\mathcal{H}_{\Gamma_{2,L}}$ on $\Gamma_{2,L}$ is provided by the spin-network states $\ket{(j_\ell),i_t, i_s}$. In the holonomy representation, they are given by:
\be
\psi_{i_{t}, i_{s}, j_{\ell}}(h_{\ell}) = \!  \scalar{h_\ell}{ (j_\ell), i_t, i_s} = \!\!\! \sum_{m_\ell,n_\ell} \prod_\ell \sqrt{2j_\ell+1} \left[ D^{j_\ell}(h_\ell)\right]_{m_\ell n_\ell} \! \left[i_s\right]^{m_1 m_2 m_3 m_4} \! \left[i_t\right]^{n_1 n_2 n_3 n_4} \, . \nonumber
\ee
A generic state can be expanded as:
\be
\ket{\Psi_{\Gamma_{2,L}}} = \sum_{j_\ell, i_{t,s}} c_{j_{\ell},i_{t,s}} \, \ket{ (j_\ell),i_t, i_s} \, ,
\ee
where $j_\ell = 1/2,1,\dots $, and the sum over intertwiner indices runs over a finite set of states $i_s = 1, \dots, \dim \mathcal{H}_s$, where $\mathcal{H}_s$ is the intertwiner space at the source node, and similarly for the target.

The automorphism group of $\Gamma_{2,L}$ is the direct product of two symmetric groups, $\mathfrak{S}_{N,L} \equiv \mathfrak{S}_{N} \times \mathfrak{S}_{L}$, describing the permutations of the nodes and the links, respectively, with $N=2$. The total number of automorphisms is $|\mathfrak{S}_{2, L} | = 2! L!$. An automorphism invariant state $\ket{\Psi_{\Gamma_{2,L}}}$ in $\mathcal{K}_{\Gamma_{2,L}}$ satisfies:
\be
 U_{\mathrm{A}} \ket{\Psi_{\Gamma_{2,L}}} = \ket{\Psi_{\Gamma_{2,L}}}, \quad \forall \, \mathrm{A} \in \mathfrak{S}_{2,L} \, .
 \ee
The density matrix representation $\rho_{\Gamma_{2,L}}$ of $\ket{\Psi_{\Gamma_{2,L}}}$ is also invariant under $\mathfrak{S}_{2,L}$:
\be
U_{\mathrm{A}} \, \rho_{\Gamma_{2,L}} \,  U_{\mathrm{A}}^{-1} = \rho_{\Gamma_{2,L}} \,.
\ee

\paragraph*{Invariant observables on $\Gamma_{2,L}$.} 

The physical observable $\mathcal{O}$ associated with a gauge-fixed observable $\tilde{\mathcal{O}}$ on $\Gamma_{2,L}$ is the operator:
\be
\mathcal{O}_{\mathrm{inv}} = \frac{1}{2! L!} \sum_{\mathrm{A} \in \mathrm{Aut}(\Gamma_{2,L})} \!\! U_{\mathrm{A}} \, \mathcal{O} \, U_{\mathrm{A}}^{-1} \, ,
\label{eq:invariant-observables-dipole}
\ee
which is invariant under $\mathfrak{S}_{2,L}$. The automorphism-invariant space $\mathcal{K}_{\Gamma_{2,L}}$ is preserved by the action of $\mathcal{O}_{\mathrm{inv}}$, so that the restriction $\mathcal{O}_{\mathrm{inv}}: \mathcal{K}_{\Gamma_{2,L}} \to \mathcal{K}_{\Gamma_{2,L}}$ is well defined. The expectation values of the operators $\mathcal{O}_{\mathrm{inv}}$ and $\mathcal{O}$ are the same for an invariant state $\rho_{\Gamma_{2,L}}$,
\be
\mathrm{Tr}(\rho_{\Gamma_{2,L}} \, \mathcal{O}_{\mathrm{inv}}) = \mathrm{Tr}(\rho_{\Gamma_{2,L}} \, \mathcal{O}) \, .
\ee

If $\Sigma$ is a subgraph of $\Gamma_{2,L}$ with node set $\mathcal{N}(\Sigma)$, we say that an operator $\mathcal{O}_{\Sigma}: \mathcal{H}_{\Gamma_{2,L}} \to \mathcal{H}_{\Gamma_{2,L}}$ is a local one-subgraph operator at the subgraph $\Sigma$ if its action on the spin-network basis on $\Gamma_{2,L}$ has the form:
\be
\mathcal{O}_{\Sigma} \ket{ (j_\ell),i_t, i_s} = \!\!\!\! \sum_{\hspace{0.3cm} \{i_{\bar{n}},(j_{\bar{\ell}})\}_{\Sigma}} \!\!\!\mathcal{O}_{\Sigma, (j_\ell),i_t,i_s; \{i_{\bar{n}},(j_{\bar{\ell}})\}_{\Sigma}} \, \ket{ \{ i_{\bar{n}}, (j_{\bar{\ell}}) \}_{\Sigma}, \{i_{n^{c}}, (j_{\ell^{c}})\}_{\Sigma^c}}\, ,
\ee
where the intertwiner indices and spins $\{i_{\bar{n}},(j_{\bar{\ell}})\}$ and $\{i_{n^{c}}, (j_{\ell^{c}})\}$ are assigned to the nodes and links of the subgraphs $\Sigma$ and its complement $\Sigma^c$, respectively. The one-subgraph operator $\mathcal{O}_{\Sigma}$ has a non-trivial action only on the degrees of freedom $\{i_{\bar{n}},(j_{\bar{\ell}}) \}_{\Sigma}$ in $\Sigma$.

Under an automorphism A for which $\mathrm{A}(\Sigma)=\Sigma'$, the operator $\mathcal{O}_{\Sigma}$ transforms as:
\begin{align}
U_{\mathrm{A}} \, \mathcal{O}_{\Sigma} \, U_{\mathrm{A}}^{-1} \ket{ (j_\ell),i_t, i_s} & =  (-1)^R \, U_{\mathrm{A}} \, \mathcal{O}_{\Sigma} \ket{ (j'_\ell),i'_t, i'_s} 
\nonumber \\
	&= (-1)^R \, U_{\mathrm{A}}  \!\!\!\! \sum_{\hspace{0.3cm} \{i_{\bar{n}}, (j_{\bar{\ell}})\}_{\Sigma}} \!\!\! \mathcal{O}'_{\Sigma, (j_\ell'),i_t',i_s'; \{i_{\bar{n}}', (j_{\bar{\ell}}')\}_{\Sigma}} \\
	& \qquad \qquad \qquad \qquad \quad \times \ket{\{i_{\bar{n}}',(j_{\bar{\ell}}')\}_{\Sigma}, \{i'_{n^{c}}, (j'_{\ell^{c}})\}_{\Sigma^c}}  
\nonumber \\
	&=\!\!\!\! \sum_{\hspace{0.3cm} \{i_{\bar{n}},(j_{\bar{\ell}})\}_{\Sigma'}} \!\!\!\mathcal{O}_{\Sigma', (j_\ell),i_t,i_s;\{i_{\bar{n}},(j_{\bar{\ell}})\}_{\Sigma'}} \, \ket{\{i_{\bar{n}},(j_{\bar{\ell}})\}_{\Sigma'}, \{i_{n^{c}},(j_{\ell^{c}})\}_{\Sigma'^{c}}} 
\nonumber \\
	&= \mathcal{O}_{\Sigma'} \ket{(j_{\ell}), i_t, i_s} \,,
	\label{eq:transformation-observables-dipole1}
\end{align}
where the action~(\ref{eq:autoaction}) of an automorphism on the spin-network states (with $A \to A^{-1}$ and $U_A \to U^{-1}_A$) was used, the matrix $\mathcal{O}'_{\Sigma}$ describes the operator $\mathcal{O}_{\Sigma}$ in the basis $\ket{ (j'_\ell),i'_t, i'_s}$, and the third line consists of the definition of the matrix elements of $\mathcal{O}_{\Sigma'}$. The original local operator at the subgraph $\Sigma$ is thus transformed into a local operator at the subgraph $\Sigma'$:
\be
U_{\mathrm{A}} \, \mathcal{O}_{\Sigma} \, U_{\mathrm{A}^{-1}} = \mathcal{O}_{\Sigma'} \,.
\label{eq:transformation-observables-dipole2}
\ee

From the transformation of local operators under automorphisms~(\ref{eq:transformation-observables-dipole2}) and the definition of the invariant observables~(\ref{eq:invariant-observables-dipole}), it is evident that the invariant observables $\mathcal{O}_{\Sigma, \mathrm{inv}}$ and $\mathcal{O}_{\Sigma', \mathrm{inv}}$ associated with a local operator $\mathcal{O}_{\Sigma}$ and its image $\mathcal{O}_{\Sigma'}$ under an automorphism are identical:
\be
 \mathcal{O}_{\Sigma, \mathrm{inv}} =  \mathcal{O}_{\Sigma', \mathrm{inv}} \, ,
\ee
as the expectation values of $\mathcal{O}_{\Sigma}$ and $\mathcal{O}_{\Sigma'}$ are the same for any invariant state $\rho_{\Gamma_{2,L}}$:
\be
\Tr (\rho_{\Gamma_{2,L}} \, \mathcal{O}_{\Sigma, \mathrm{inv}}) = \Tr (\rho_{\Gamma_{2,L}} \, \mathcal{O}_{\Sigma', \mathrm{inv}}) \, .
\label{eq:local-to-inv}
\ee

On a dipole graph, for any pair of isomorphic subgraphs $\Sigma,\Sigma'$, and for every isomorphism $\bar{\mathrm{A}}: \Sigma \to \Sigma'$, there exists an automorphism A of $\Gamma_{2,L}$ such that $\mathrm{A}|_{\Sigma} = \bar{\mathrm{A}}$ and $\mathrm{A}^{-1}|_{\Sigma'} = \bar{\mathrm{A}}^{-1}$. From such a graph homogeneity of $\Gamma_{2,L}$, the space of invariant observables on a subgraph $\Sigma$ covers the entire space of invariant observables on subgraphs that are isomorphic to $\Sigma'$ in $\Gamma_{2,L}$. We refer to the invariant operator associated with a local one-subgraph operator $\mathcal{O}_{\Sigma}$ as an invariant one-subgraph operator on $\Gamma_{2,L}$.

A basic set of invariant observables is associated with the smallest subgraphs, as for instance the volume of a node $(\centering\includegraphics[trim=0 -0.05cm 0 0cm, scale=0.65]{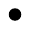})$, the area at a link $({\centering \includegraphics[trim=0 -0.15cm 0 0cm, scale=0.65]{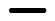}})$, and the dihedral angle at a wedge $(\centering \includegraphics[trim=0 0.1cm 0 0cm, scale=0.65]{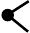})$. We define a wedge as an ordered pair of links at a node. Invariant observables associated with local one-node, one-link and one-wedge observables on $\Gamma_{2,L}$ can in general be defined as:
\be
\mathcal{O}^{(\includegraphics[scale=0.65]{node.pdf})}_{\Gamma_{2,L}} = \frac{1}{2} \sum_{n} \mathcal{O}_n\,, \quad  \mathcal{O}^{(\includegraphics[trim=0 -0.06cm 0 0cm, scale=0.65]{link.pdf})}_{\Gamma_{2,L}} = \frac{1}{2L} \sum_{\ell} \mathcal{O}_{\ell}\,, \quad \mathcal{O}^{(\centering\includegraphics[trim=0 0.2cm 0 0cm, scale=0.65]{wedge.pdf})}_{\Gamma_{2,L}} = \frac{1}{2L(L-1)} \sum_w \mathcal{O}_w \,,
\label{eq:one-nlw}
\ee
for local observables $\mathcal{O}_n$, $\mathcal{O}_\ell$ and $\mathcal{O}_w$ invariant under automorphisms that preserve the node $n$, the link $\ell$ or the wedge $w$, respectively. The observables $\mathcal{O}^{(\includegraphics[scale=0.65]{node.pdf})}_{\Gamma_{2,L}}$, $\mathcal{O}^{(\includegraphics[trim=0 -0.06cm 0 0cm, scale=0.65]{link.pdf})}_{\Gamma_{2,L}}$ and $\mathcal{O}^{(\centering\includegraphics[trim=0 0.2cm 0 0cm, scale=0.65]{wedge.pdf})}_{\Gamma_{2,L}}$ represent the averages of the local operators over the graph. From Eq.~(\ref{eq:local-to-inv}), for any invariant state $\rho_{\Gamma_{2,L}}$, the expectation values of the invariant operators $\mathcal{O}^{(\includegraphics[scale=0.65]{node.pdf})}_{\Gamma_{2,L}}$, $\mathcal{O}^{(\includegraphics[trim=0 -0.06cm 0 0cm, scale=0.65]{link.pdf})}_{\Gamma_{2,L}}$ and $\mathcal{O}^{(\centering\includegraphics[trim=0 0.2cm 0 0cm, scale=0.65]{wedge.pdf})}_{\Gamma_{2,L}}$ can be computed at a fixed local region:
\be
\Tr (\rho_{\Gamma_{2,L}} \mathcal{O}^{(\alpha)}_{\Gamma_{2,L}}) = \Tr (\rho_{\Gamma_{2,L}} \mathcal{O}_i)\,, \quad (\alpha, i) = \{(\includegraphics[scale=0.65]{node.pdf}, n), (\includegraphics[trim=0 -0.06cm 0 0cm, scale=0.65]{link.pdf}, \ell), (\centering\includegraphics[trim=0 0.2cm 0 0cm, scale=0.65]{wedge.pdf}, w)\} \,.
\label{eq:average-invariant}
\ee

As discussed in Section \ref{sec:2.1}, the local intrinsic geometry at a node is described by the Penrose metric operator, given as a gauge-fixed operator in Eq.~(\ref{eq:shapeoperator}). The invariant observable associated with the Penrose metric operator at a fixed node, or of any function of it, can be easily constructed following the general averaging introduced in Eq.~(\ref{eq:invariant-observables-dipole}). The volume, face areas and angles $\{V_n, A_{na}, \cos \theta_{ab}(n)\}$ are local one-node, one-link and one-wedge observables constructed from the Penrose metric, respectively. We denote the associated invariant observables by $V, A, \cos \Theta$, respectively. They read:
\begin{align}
 V &:= \frac{\kappa_0}{4 \sqrt{2}} \,\, \sum_{n} \, \sum_{a < b < c \leq L} \, \sqrt{ \big| \textstyle{\frac{i}{\mathfrak{a}_{0}}} \, [g_{ab}(n), g_{ac}(n)] \, \big|}  \,,  \nonumber \\
A &:= \frac{1}{L} \, \sum_{a} \, \sqrt{g_{aa}(n)}\,, \nonumber \\ 
\cos \Theta &:= \frac{1}{2 L(L-1)} \sum_n \, \sum_{a \neq b \leq L} \frac{g_{ab}(n)}{\sqrt{g_{aa}(n)} \sqrt{g_{bb}(n)}} \,.
\label{eq:invariant-local-observable-dipole}
\end{align}
Notice that the set of operators~\eqref{eq:invariant-local-observable-dipole} lacks any information concerning the elements of the graph $\Gamma_{2,L}$. Here, the factor for the angle includes all possible oriented wedges and, the face area at each link of the source and target nodes is equal, i.e. $A_{a} \equiv  A_{sa} = A_{ta}$.

A direct consequence of Eq.~(\ref{eq:average-invariant}) is that the expectation value of the invariant observables $\{A, V, \cos \Theta\}$ can be computed as the expectation values of $V_n, A_{na}$ and $\cos \theta_{ab}(n)$ at any chosen node, link or wedge. In particular, the average $\langle \cos \Theta \rangle_{\Psi_{\Gamma_{2,L}}}$ and dispersion $(\Delta\cos \Theta)_{\Psi_{\Gamma_{2,L}}}$ of the cosine of the dihedral angle with respect to an invariant state $\ket{\Psi_{\Gamma_{2,L}}}$ are given by:
\be
\langle \cos \Theta \rangle_{\Psi_{\Gamma_{2,L}}} = \langle \cos \theta_{ab}(n) \rangle_{\Psi_{\Gamma_{2,L}}} \,, \quad \forall \, a \neq b \,,
\label{eq:average-dihedral-angle}
\ee
and
\begin{align}
(\Delta\cos \Theta)_{\Psi_{\Gamma_{2,L}}}^2 &= \frac{1}{2 L (L-1)} \,\, (\Delta\cos \theta_{ab}(n))^2_{\Psi_{\Gamma_{2,L}}} \nonumber \\
& \quad +  \frac{1}{4 L^2 (L-1)^2} \, {2 L(L-1) \choose 2} \, \langle  \cos \theta_{ab,ac}(n)  \rangle_{\Psi_{\Gamma_{2,L}}} \,, \quad \forall \, a \neq b \neq c \,,
\label{eq:dispersion-dihedral-angle}
\end{align}
where the dispersion $(\Delta\cos \theta_{ab}(n))^2_{\Psi_{\Gamma_{2,L}}}$ for any wedge $ab$ is
\be
(\Delta\cos \theta_{ab}(n))_{\Psi_{\Gamma_{2,L}}} = \sqrt{\langle \cos^2 \theta_{ab}(n) \rangle_{\Psi_{\Gamma_{2,L}}}- \langle \cos \theta_{ab}(n) \rangle_{\Psi_{\Gamma_{2,L}}}^2}
\ee
and the connected correlation function $ \langle  \cos \theta_{ab,ac}(n)  \rangle_{\Psi_{\Gamma_{2,L}}}$ for any pair of wedges $ab$ and $ac$ is given by
\be
 \langle  \cos \theta_{ab,ac}(n)  \rangle_{\Psi_{\Gamma_{2,L}}} = \langle \cos \theta_{ab}(n) \cos \theta_{ac}(n) \rangle_{\Psi_{\Gamma_{2,L}}} -  \langle \cos \theta_{ab}(n) \rangle_{\Psi_{\Gamma_{2,L}}}  \langle \cos \theta_{ac}(n) \rangle_{\Psi_{\Gamma_{2,L}}} \,.
\ee
Similar formulas hold for the area and volume.

At this point, we have all the necessary tools to extract local properties of the quantum geometry on a dipole graph.  We will focus on the determination of expected values and dispersions of the local observables described in Eq.~(\ref{eq:invariant-local-observable-dipole}).

\subsection{Bell-network states on $\Gamma_{2,L}$ with $L=4$}

The Bell-network state $\ket{\Gamma_{2,L},\mathcal{B},\lambda}$ on a dipole graph $\Gamma_{2,L}$ reads~\cite{bell}:
\begin{equation}
\ket{\Gamma_{2},\mathcal{B},\lambda}= \frac{1}{\sqrt{\mathcal{N}_{\lambda}}}\,\sum_{j_{\ell}, k} \big( \prod_\ell (-\lambda)^{2 j_\ell} \big)  \; \ket{ (j_{\ell}), i_{k}, (\zeta i)_{k}} \,,
\label{eq:bell-network-dipole}
\end{equation}
where $\mathcal{N}_{\lambda}$ is a normalization factor. The action of the antilinear map $\zeta:\mathcal{V}_{j} \to \mathcal{V}_{j}$ in Eq.~(\ref{eq:bell-network-dipole}) is given by:
\be
(\zeta v)^m = (\bar{v})^{-m} (-1)^{j-m} \,.
\ee
and corresponds to the operation of time-reversal for spin states in $\mathcal{V}_{j}$.

The invariance of any observable of the intrinsic geometry under the time-reversal operation $\zeta$ follows from the invariance of $g_{ab}$ under the action of $\zeta$. Consequently, the intertwiners on the source and target nodes $(s,t)$ describe the same local intrinsic geometry. Expressing the state $\ket{ (j_{\ell}), i_{k}, (\zeta i)_{k}}$ in the magnetic number basis, we find:
\be
\ket{ (j_{\ell}), i_{k}, (\zeta i)_{k}} = \!\!\!\!\! \sum_{\tilde{m}_\ell, m_\ell = -j_\ell}^{j_\ell} \! (i_{k})^{\tilde{m}_1 \cdots \tilde{m}_L} \, (\bar{i}_{k})^{-m_1 \cdots -m_L} \bigotimes_{\ell=1}^{L} \,(-1)^{j_\ell - m_\ell} \, \ket{j_\ell, \tilde{m}_\ell}_s \otimes \ket{j_\ell, m_\ell}_t \,. \nonumber
\ee
The state exhibits a perfect correlation such that if the measurement of an observable of the geometry of the quantum polyhedron $s$ has an outcome, then the observation of the same quantity at the quantum polyhedron $t$ gives the same result.

The Bell-network state~(\ref{eq:bell-network-dipole}) on $\Gamma_{2,L}$ is a CBN state, invariant under automorphisms. A state with fixed spins $j_\ell$ can be obtained by projecting into the invariant subspace $\mathcal{K}_{\Gamma_{2, j_\ell}}$. An invariant state can then be obtained by averaging over the action of all automorphisms. An explicit expression for the invariant CBN state on $\Gamma_{2,4}$ with definite spins $j_\ell$ is obtained from Eq.~(\ref{eq:gamma-fixed-j}):
\be
\ket{\Gamma_{2,4}, \mathcal{B},\{j_\ell\}}= \frac{1}{4!} \, \sum_{\sigma} \, \frac{1}{\sqrt{\dim \mathcal{H}_n(j_\ell)}} \!\! \sum_{k=1}^{\dim \mathcal{H}_n(j_\ell)} \! \ket{ (j_{\sigma(\ell)}), i_{k}, (\zeta i)_{k}} \,,
\label{eq:gamma-fixed-j-L=4}
\ee
where the dimension of the intertwiner space $\dim \mathcal{H}_n(j_\ell)$,
\begin{equation}
\dim \mathcal{H}_n(j_\ell) =  \mathrm{min}(j_a+j_b, j_c+j_d) -  \mathrm{max}(|j_a-j_b|,|j_c-j_d|) + 1 \,,
\label{eq:}
\end{equation}
where $\sigma$ represents a permutation of the links, and the dimension of the intertwiner space,
\begin{equation}
\dim \mathcal{H}_n(j_\ell) =  \mathrm{min}(j_a+j_b, j_c+j_d) -  \mathrm{max}(|j_a-j_b|,|j_c-j_d|) + 1 \,,
\end{equation}
is invariant under the action of permutations $\sigma$ of $\mathfrak{S}_L$.

In our analysis for the effective geometry on $\Gamma_{2,4}$, we choose our invariant state to be the CBN state~(\ref{eq:gamma-fixed-j-L=4}) with definite spins $j_\ell$.

\subsection{Effective geometry as spherical tetrahedron} 
\label{sec:effec-spherical}

As discussed in section~\ref{subsec:dipole-graph}, the local properties of quantum geometry on $\Gamma_{2,4}$ are characterized by a set of invariant operators $V$, $A$, and $\cos\Theta$~(\ref{eq:invariant-local-observable-dipole}), describing the volume, area and angle operators associated with any given set $\{n, \ell, ab\}$. The effective geometry at a node is that described by the expectation values of $V$, $A$, and $\cos\Theta$. In this section, we numerically compute such expectation values and associated dispersions for CBN states at fixed spins.

The operator $\cos \Theta$ represents the measurement of the cosine of the dihedral angle between adjacent faces. Every pair of links on $\Gamma_{2,4}$ represents two adjacent boundary faces. Let $D_{ab}$ denote the expectation value of the corresponding dihedral angle operator, i.e. $D_{ab} \equiv \langle \cos \theta_{ab} \rangle$. The expectation value of $\cos\Theta$ with respect to the CBN state with fixed spins~(\ref{eq:gamma-fixed-j-L=4}) is given by:
\begin{align}
D_{\mathcal{B}}(j_\ell) := \langle \cos\Theta \rangle_{\mathcal{B},j_\ell} = \langle \cos \theta_{ab}(n) \rangle_{\mathcal{B},j_\ell} = \frac{1}{4 !} \sum_\sigma \, D_{ab}(j_{\sigma({\ell})}) = \frac{1}{6} \, \sum_{a < b} \, D_{ab}(j_\ell) \,.
\end{align}
Here, we used the equivalence~(\ref{eq:average-dihedral-angle}) of the expectation values of $\cos \Theta$ and $\cos \theta_{ab}(n)$, the following symmetry relations for $D_{ab}$:
\be
D_{ab}(j_{\sigma({\ell})}) = D_{\sigma^{-1}(a) \sigma^{-1}(b)}(j_\ell) \,,
\label{eq:symmetry-dihedral}
\ee
and, in particular, the invariance of the operator under spin permutations at the wedge:
\be
D_{ab}=D_{ba} \, .
\label{eq:permutation-dihedral}
\ee

The expectation value of $\cos \theta_{ab}(n)$ for the Bell-network state at a fixed spin configuration $(j_\ell)$ is given by:
\begin{align}
D_{ab}(j_\ell) = \! \frac{1}{\dim \mathcal{H}_n(j_\ell)}  \sum_{k=k_{ab}^{\mathrm{min}}}^{k_{ab}^{\mathrm{max}}} \!\! \chi_{ab}(k,j_\ell)  \frac{\big(k(k+1) - j_a(j_a+1) - j_b(j_b+1)\big)}{2 \sqrt{j_a(j_a+1)j_b(j_b+1)}}\,,
\end{align}
where $\dim \mathcal{H}_n(j_\ell) = k_{ab}^{\mathrm{max}}  - k_{ab}^{\mathrm{min}} + 1$ with $k_{ab}^{\mathrm{min}} = \mathrm{max}(|j_a-j_b|,|j_c-j_d|)$ and $k_{ab}^{\mathrm{max}} = \mathrm{min}(j_a+j_b, j_c+j_d)$. The characteristic function $\chi_{ab}(k, j_\ell)$ implements the Clebsch-Gordon conditions:  
\begin{align}
\chi_{ab}(k, j_\ell) = \begin{cases}  
   1 & \mathrm{if}\,\, |j_a-j_b| \leq k \leq |j_a+j_b| \,\,\, \mathrm{and} \,\,\, |j_c-j_d| \leq k \leq |j_c+j_d| \,\,\, \mathrm{with} \\
   & \quad  j_a+j_b+k, \,\, j_c+j_d+k \in \mathbb{Z}^+ \\
     0 & \mathrm{otherwise} \,. \nonumber
     \end{cases}
\end{align}

Explicit formulas for $D_{\mathcal{B}}(j_\ell)$ can be obtained for several special sets of spin values:
\begin{enumerate}
\item[\textit{i.}] When all spins are equal, $j_1=j_2=j_3=j_4 \equiv j$,
\begin{align}
D^{(1)}_{\mathcal{B}}(j) = - \frac{1}{3}
\label{eq:D-j}
\end{align}
corresponds to the cosine of the dihedral angle of a \textit{flat regular tetrahedron}. The flat regular tetrahedron is then the only solution for the case of equal spins. 
\item[\textit{ii.}]  When three spins are equal, $j_1=j_2 = j_3 \equiv j$ and $j_4 \equiv j'$,
\begin{align}
D^{(2)}_{\mathcal{B}}(j, j') = -\frac{1}{4} -\frac{1}{6} \sqrt{\frac{j'(j'+1)}{j(j+1)}} + \frac{1}{12} \frac{j'(j'+1)}{j(j+1)} \,,
\label{eq:D-13}
\end{align}
with $j' \leq j$ or $j \leq j' \leq 3 j$, so that the Clebsch-Gordon conditions are satisfied.
\item[\textit{iii.}] For two pairs of equal spins, $j_1=j_2 \equiv j$ and $j_3=j_4 \equiv j' \leq j$,
\begin{align}
D^{(3)}_{\mathcal{B}}(j,j') = -\frac{1}{9}\bigg(2+ 2 \sqrt{\frac{j'(j'+1)}{j(j+1)}} - \frac{j'(j'+1)}{j(j+1)}\bigg) \,.
\label{eq:D3}
\end{align}
\item[\textit{iv.}] When two spins are equal, $j_3 = j_4 \equiv j$: 
\begin{align}
D^{(4)}_{\mathcal{B}}(j,j_1,j_2) &= -\frac{1}{36} \Bigg(2 \sqrt{\frac{j_1(j_1+1)}{j_2(j_2+1)}} + 4 \sqrt{\frac{j_1(j_1+1)}{j(j+1)}} - 2 \frac{\big(j_1(j_1+1)-3 j_2(j_2+1)\big)}{\sqrt{j_2(j_2+1)j(j+1)}} \nonumber \\
& - \frac{j_1(j_1+1) + 3\big(j_2(j_2+1) - 2 j(j+1)\big)}{j(j+1)}\Bigg)\,,
\label{eq:D4}
\end{align}
with  $j_1 \leq j_2 \leq j$.
\end{enumerate}

For generic spins, a closed form for $D_{\mathcal{B}}(j_\ell)$ can be obtained by computing $D_{12}(j_\ell)$ and using the symmetry relations~(\ref{eq:symmetry-dihedral}):
\begin{align}
D_{\mathcal{B}}(j_\ell) &= \frac{1}{6} \, \big(D_{12}(j_1,j_2,j_3,j_4) + D_{12}(j_1,j_3,j_2,j_4) + D_{12}(j_1,j_4,j_2,j_3) \nonumber \\
& + D_{12}(j_2,j_3,j_1,j_4) + D_{12}(j_2,j_4,j_1,j_3) + D_{12}(j_3,j_4,j_1,j_2) \big)\,,
\label{eq:generic-D}
\end{align}

This formula will be employed in numerical calculations later in the paper.

\begin{figure}
    \centering
    \includegraphics[width=7.5cm]{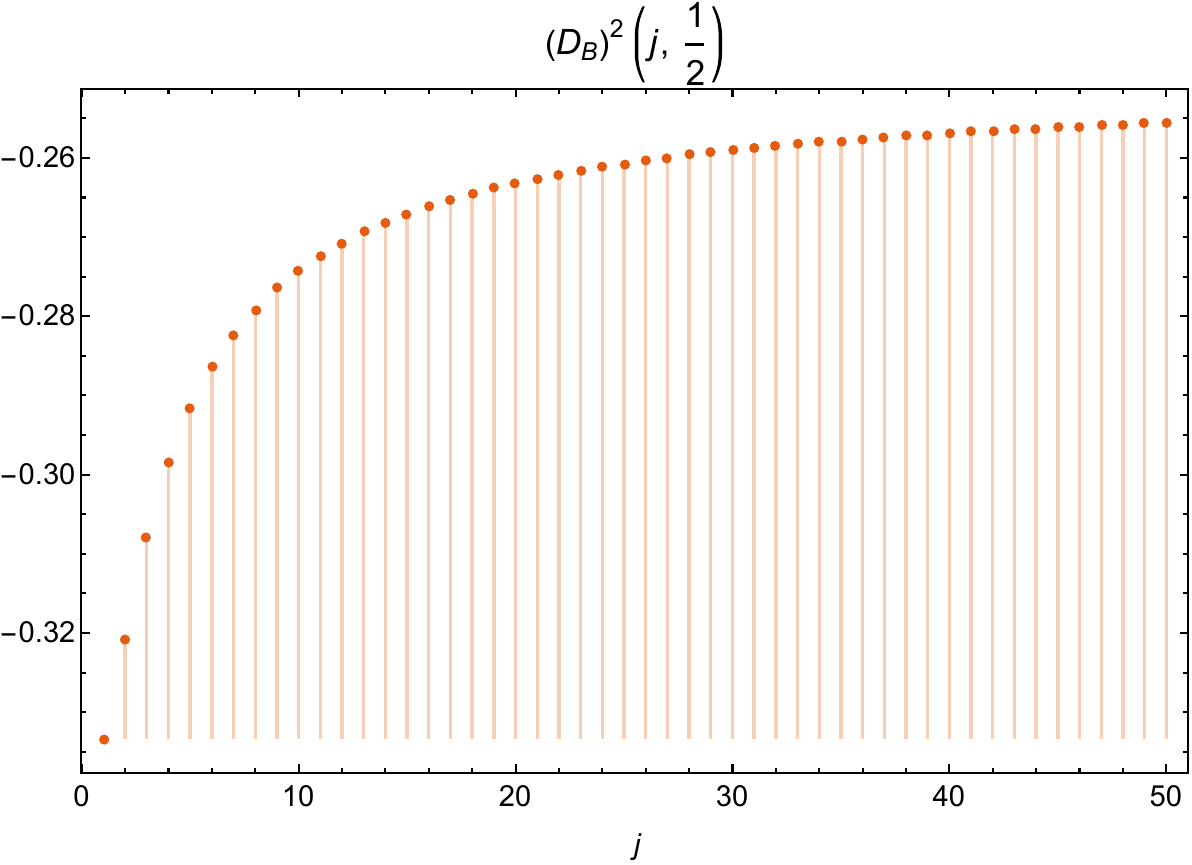}
    \caption{Expectation value $D^{(2)}_{\mathcal{B}}(j, j')$ of the cosine of the dihedral angle operator $\cos\Theta$ for CBN states $\ket{\Gamma_{2,4}, \mathcal{B},j_\ell}$ with $j_1=j_2=j_3 = j$ and $j_4 = j'=1/2$.}
    \label{fig:D_2}
\end{figure}

Note that $D_{\mathcal{B}}(j_\ell)$ is completely symmetric in the spins. It describes the average of the cosine of the dihedral angle as measured in any pair of faces at one of the nodes. Due to isotropy, distinct dihedral angles $D_{ab}$ cannot be distinguished, and the geometry at the node is in this sense regular, in spite of the fact that the state displays fluctuations over anisotropic geometries, characterized by the spins $j_\ell$. However, the average geometry in general does not describe a regular flat tetrahedron, for which one must have $D_{\mathcal{B}}=-1/3$. This can be seen from the Eqs.~(\ref{eq:D-j})--(\ref{eq:D4}). If all spins are equal, the geometry is that of a flat regular tetrahedron, as described by Eq.~(\ref{eq:D-j}). On the other hand, in the cases (ii) and (ii), for which $D_{\mathcal{B}}$ is given by Eqs.~(\ref{eq:D-13}) and (\ref{eq:D3}), respectively, there is no solution for  $D_{\mathcal{B}}=-1/3$, except for $j = j'$, which reduces to the case of all spins equal. In~Fig.~\ref{fig:D_2}, we plot $D^{(2)}_{\mathcal{B}}(j, j')$ with $j'=1/2$ as a function of $j$. As $j \to 1/2$, the average $D_{\mathcal{B}}$ approaches the value $-1/3$ that characterizes a regular flat tetrahedron. For larger $j$, the average progressively increases, moving far from the flat limit, consistently with the fact that the only solution of $D_{\mathcal{B}}^{(2)}(j,j')=-1/3$ is that with all spins equal.

How can the average geometry be interpreted? As the node represents a region of space bounded by four faces, such that any pair of faces are indistiguishable, it is natural to try and interpret it as a curved regular tetrahedron. The geometry of curved tetrahedra is briefly reviewed in the Appendix~\ref{sec:appendix}. In general, a curved tetrahedra is a region of a space of constant curvature bounded by four faces that are flatly embedded surfaces intersecting only at their boundaries. Its four vertices are connected by edges that are geodesic curves.

The interior of a curved tetrahedron is a region of a hyperbolic, spherical or flat space with constant curvature. The sign of the curvature can be determined from the cosines of the dihedral angles. This can be achieved by computing the determinant of the Gram matrix $G$ (see Eq.~(\ref{eq:Grammatrix})) associated with the tetrahedron. The space is hyperbolic, flat or spherical according to whether $\det G$ is negative, zero or positive, respectively. For a regular tetrahedron $T_{j_\ell}$ representing the effective geometry of a CBN state with fixed spins $j_\ell$, the Gram matrix has the special form:
\be
(G_{T_{j_\ell}})_{\alpha \beta} = \left\{ 
\begin{array}{lc}
D_{\mathcal{B}}(j_\ell) \, , & \alpha \neq \beta \, , \\
1\,, & \alpha = \beta \, .
\end{array}
\right.
\ee
Its determinant reads:
\be
\det G_{T_{j_\ell}} = (1- D_{\mathcal{B}}(j_\ell))^3 \, (1 + 3 D_{\mathcal{B}}(j_\ell))\,,
\ee
where the domain is restricted to $|D_{\mathcal{B}}(j_\ell)| < 1$. We see that a unique root for $\det G_{T_{j_\ell}}=0$ exists at $D_{\mathcal{B}}(j_\ell) = -1/3$, corresponding to the case of a flat regular tetrahedron. If we consider $D_{\mathcal{B}}(j_\ell)$ as a free parameter, independently of any specific constraints imposed by the set of spins $j_\ell$, then the solutions for $D_{\mathcal{B}}(j_\ell) < -1/3$ and $D_{\mathcal{B}}(j_\ell) > -1/3$ correspond to hyperbolic and spherical tetrahedra, respectively. The value of $D_{\mathcal{B}}(j_\ell)$ is determined by the spins, however, and we need to investigate what is the range of values that it can assume for generic spins. Let us consider some special cases.

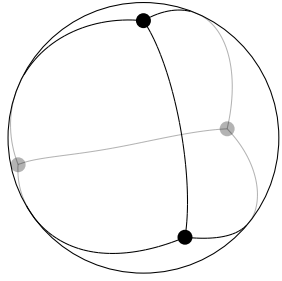
\begin{figure}
\begin{centering}
\pgfmathsetmacro{\RadiusSphere}{3}
\tdplotsetmaincoords{60}{131}
\begin{tikzpicture}[scale=0.6,tdplot_main_coords]
\hspace{0.5cm}
  \coordinate (O) at (0,0,0);
 \draw[tdplot_screen_coords,fill opacity=0.3] (0,0,0) circle (3);
 \path (xyz spherical cs:radius=\RadiusSphere,latitude=90,longitude=00)
  node[fill,circle,inner sep=2pt]{}; 
 \foreach \X in {0,1,2} 
 {\draw plot[variable=\t,domain=0:0-109.5,smooth] 
  (z spherical cs:radius=\RadiusSphere,theta=\t,phi=\X*120) 
  \ifnum\X=2
  node[fill,circle,inner sep=2pt]{}
  \else
  node[fill,circle,inner sep=2pt,opacity=0.3]{}
  \fi;}
 \draw plot[variable=\t,domain=0:360,smooth,samples=121] 
 (z spherical cs:radius=\RadiusSphere,theta={-109.5-14*abs(sin(1.5*\t))},phi=\t);
\end{tikzpicture}
\end{centering}
\caption{A triangulation of $2$-sphere induced by a spherical tetrahedron.}
\label{fig:sphericalt}
\end{figure}

If all spins are equal, $j_\ell=j$, then we have $D_{\mathcal{B}}(j_\ell) = D^{(1)}_{\mathcal{B}}(j) = - 1/3$, and $\det G_{T_{j_\ell}} =0$, i.e. a flat tetrahedron. When three spins are equal, $j_1=j_2=j_3=j$ and $j_4=j'$, we have $D_{\mathcal{B}}(j_\ell)= D^{(2)}_{\mathcal{B}}(j,j')$, given in explicit form in eq.~(\ref{eq:D-13}). For any fixed $j$, the function $D^{(2)}_{\mathcal{B}}(j,j')$ has a unique global minimum of $-1/3$ at $j'=j$, when it reduces to the case of all spins equal. Similarly, for the case of two pairs of equal spins, $j_1=j_2=j$ and $j_3=j_4=j'$, in which case $D_{\mathcal{B}}(j_\ell)= D^{(3)}_{\mathcal{B}}(j,j')$ is given by eq.~(\ref{eq:D3}), again there is a unique global minimum of $-1/3$ at $j'=j$. In addition, we checked numerically that with two equal spins $j_3=j_4=j$, the function $D_{\mathcal{B}}(j_\ell)= D^{(4)}_{\mathcal{B}}(j,j')$ given in eq.~(\ref{eq:D4}) also achieves a minimum when all spins are equal, and that the same is true in the general case of spins $j_1\leq j_2 \leq j_3 \leq j_4$. As an evidence that no other minima can be found at larger spins, we also note that, considering an interpolation of $D_{\mathcal{B}}(j_\ell)$ with continuous spins, the extrema of $D_{\mathcal{B}}(j_\ell)$ only occur at $j_\ell=j$. In short, except when all spins are equal, the Gram matrix is that of a spherical tetrahedron.

Therefore, for a CBN state $\ket{\Gamma_{2,4}, \mathcal{B}, \{j_\ell\}}$ on a dipole graph $\Gamma_{2,4}$, compelling evidence suggests that the expected average geometry at each node is that of a \textit{flat regular tetrahedron} $T_{\mathbb{R}^3}$ exclusively when all spins are equal, $j_\ell = j$. For all other configurations, the average dihedral angles are those of a \textit{regular spherical tetrahedron} $T_{\mathbb{S}^3}$ (see Fig~\ref{fig:sphericalt}).

\begin{figure}
    \centering
 \includegraphics[width=7.25cm]{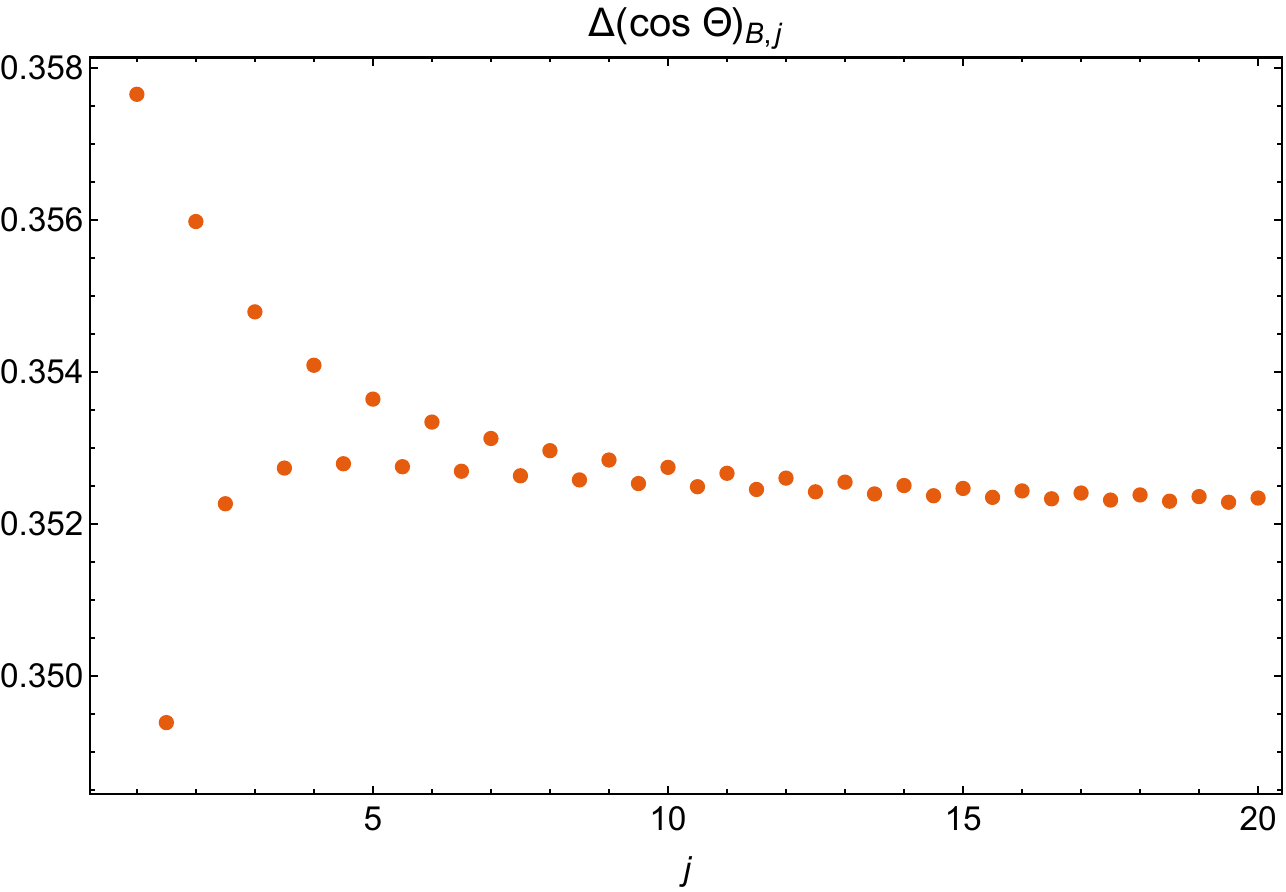}%
    \caption{Dispersion $\Delta (\cos \Theta)_{\mathcal{B},j}$ of the cosine of the dihedral angle operator $\cos\Theta$ for CBN states $\ket{\Gamma_{2,4}, \mathcal{B},\{j_\ell\}}$ with $j_\ell = j$.}%
    \label{fig:dispersion-costheta}%
\end{figure}

We now consider the dispersion of $\cos \Theta$ around its expectation value. The dispersion $\Delta (\cos \Theta)_{\mathcal{B},j_\ell}$, defined in eq.~\eqref{eq:dispersion-dihedral-angle}, for equal spins $j_\ell = j$ can be written as:
\begin{align}
\Delta (\cos \Theta)^2_{\mathcal{B},j}\,\big|_{L=4} &= \frac{23}{48}  \left[\frac{1}{2j+1} \, \sum_{k, l = 0}^{2j} W_{kl}(j) \, \left(\frac{l(l+1) - 2j(j+1)}{2 j(j+1)}\right)^2 -\frac{1}{9}\right]  \nonumber \\
& \quad +  \frac{1}{24} \, \left(\frac{16}{45} + \frac{1}{15 j + 15 j^2}\right) \,,
\label{eq:dispersion-dihedral-equal}
\end{align}
where $W_{kl}(j)$ is the Wigner $6j$-symbol:
\begin{align}
W_{kl}(j) = \sqrt{(2k+1) (2 l +1)} \, \begin{Bmatrix}
j & j & k \\
j & j & l
\end{Bmatrix} \,.
\end{align} 
The dispersion $\Delta (\cos \Theta)_{\mathcal{B},j}$ remain finite for large $j$ (see Fig.~\ref{fig:dispersion-costheta}), characterizing large fluctuations of the geometry at the node even in the limit, naively taken as semiclasical, of large spins.

\begin{figure}%
\begin{centering}
\includegraphics[height=5cm]{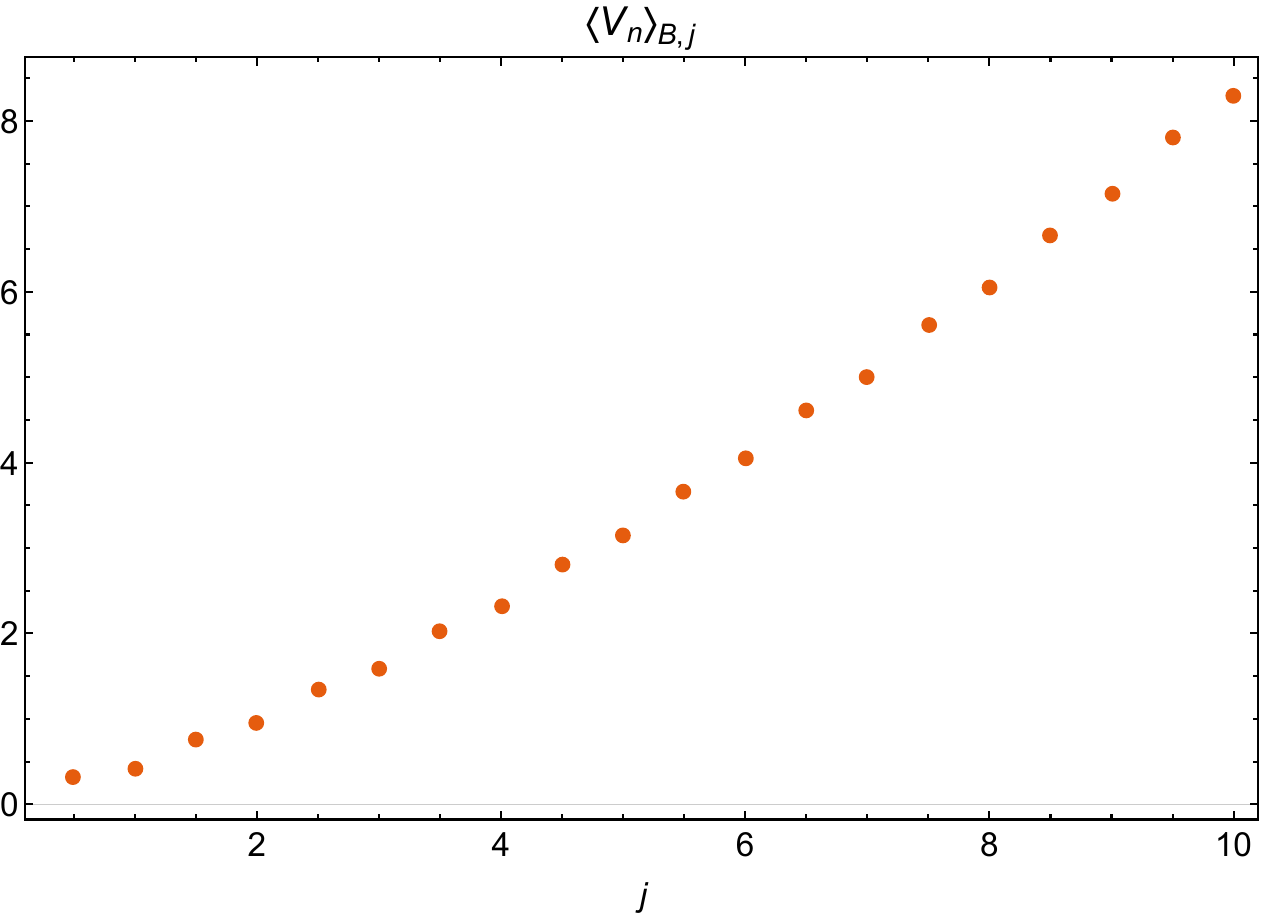}
\includegraphics[height=5cm]{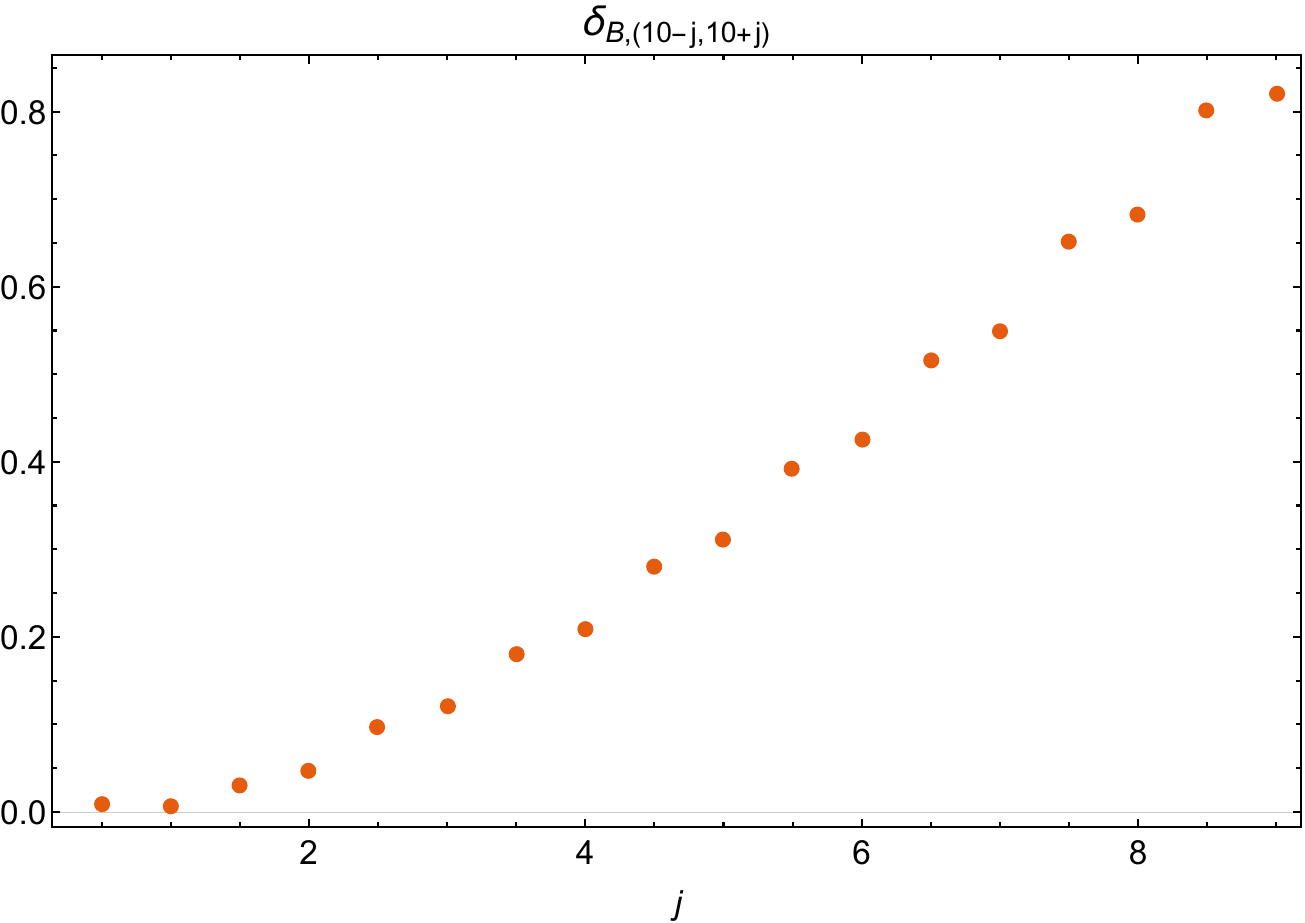}
    \caption{Expectation value of the volume operator $V$ for CBN states $\ket{\Gamma_{2,4}, \mathcal{B},\{j_\ell\}}$. \textit{Left panel:} Average volume $\langle V_n \rangle_{\mathcal{B},j}$ for equal spins $j_\ell=j$, in units of $\mathfrak{a}_{0}^{3/2}$. \textit{Right panel:} Spatial curvature witness $\delta_{\mathcal{B}}(j_\ell)$, for a total spin of $J=40$, with $j_1=j_2 = 10-j$ and $j_3 = j_4 = 10+j$.}
\label{fig:avgvolume}
\end{centering}
\end{figure}

Let us now compute the average volume. Following the equivalence~(\ref{eq:average-invariant}) of the expectation value of an invariant operator and its corresponding local operator for an invariant state, the expectation value of $V$ is equal to the expectation value of the volume operator $V_{n}$, given in eq.~(\ref{eq:volume-4-node}), at any node $n=1,2$:
\be
\langle V \rangle_{\mathcal{B},j_\ell} = \langle V_n \rangle_{\mathcal{B},j_\ell} \, .
\ee
The volume operator $V_{n}$ can be expressed in diagonal form as follows \cite{bianchilength}:
\be
V_{n}  =  \frac{\sqrt{2}}{3} \,\, \mathfrak{a}_{0}^{3/2} \, \sum_{i=1}^{\dim \mathcal{H}_n} \sqrt{|q_i(j_\ell)|} \,\, \ket{q_i} \, \bra{q_i} \,. 
\label{eq:voldiag}
\ee
where the states $\ket{q_i}$ are eigenstates of the operator $Q_n$ defined in the basis of eigenstates $\ket{k_{12}}$ of the component $g_{12}$ of the Penrose metric operator by:
\be
Q_n(j_\ell) = \sum_{k_{12}^{\mathrm{min}}+1}^{k_{12}^{\mathrm{max}}} i a(k, j_\ell) \, \big(\ket{k_{12}} \bra{(k-1)_{12}} - \ket{(k-1)_{12}} \bra{k_{12}}\big)
\ee
with
\begin{align}
a(k,j_\ell) = \frac{1}{\sqrt{4 k^2-1}} &\left[ (k+j_1+j_2+1)(-k+j_1+j_2+1)(k-j_1+j_2) \right. \nonumber \\
& \quad \times  (k+j_1-j_2) (k+j_3+j_4+1)(-k+j_3+j_4+1) \nonumber \\
& \qquad \left. \times (k-j_3+j_4)(k+j_3-j_4) \right]^{1/2} \,.
\end{align}
The eigenvalues $q_i(j_\ell)$ are real and non-degenerate. The non-zero eigenvalues appear in pairs with opposite sign. When $\dim \mathcal{H}_n$ is odd, there is a non-degenerate vanishing eigenvalue. As the eigenstates $\ket{q_i}$ provide an orthonormal basis of the intertwiner space $\dim \mathcal{H}_n$, the expectation value of $V_{n}$~(\ref{eq:voldiag}) with respect $\ket{\Gamma_{2,4}, \mathcal{B}, \{j_\ell\}}$~(\ref{eq:gamma-fixed-j}) is simply:
\be
\langle V_n \rangle_{\mathcal{B},j_\ell} = \frac{1}{\dim \mathcal{H}_n} \frac{\sqrt{2}}{3} \, \mathfrak{a}_{0}^{3/2} \, \sum_{i=1}^{\dim \mathcal{H}_n} \sqrt{|q_i(j_\ell)|}\, .
\label{eq:avgvolume}
\ee

In Fig.~\ref{fig:avgvolume}, we present numerical investigations of $\langle V_n \rangle_{\mathcal{B},j_\ell}$. We consider the cases of equal spins $j_\ell =j$ and two pairs of equal spins, $j_1=j_2$ and $j_3 = j_4$, which we have previously found to display average dihedral angles corresponding to flat and spherical tetrahedra, respectively. We first determined the average volume for the case of equal spins $j$, with a total boundary spin $J=4j$. Next, keeping a fixed total boundary spin $J$, we investigated the case of two pairs of equal spins, computing the average volume for spin configurations with $j_1=j_2 = J/4+j$ and $j_3 = j_4 = J/4-j$. The difference between the volume of the flat case (all spins equal) and those involving distinct spins provides a natural witness of spatial curvature: 
\be
\delta_{\mathcal{B}}(j_\ell) := 1 - \frac{\langle V_n \rangle_{\mathcal{B},{j_\ell}}}{\langle V_n \rangle_{\mathcal{B},J/4}}\,,
\label{eq:witcurv}
\ee
which measures the relative deviation of the average volume from its value at $j_\ell = J/4$, while maintaining a fixed total spin constraint $\sum_\ell j_\ell = J$. A specific numerical example is presented in Fig.~\ref{fig:avgvolume} to illustrate the general behavior. We see that the average value progressively deviates from that obtained for equal spins as the difference between the spins at the node increases. If the node is interpreted as a spherical tetrahedron, this indicates an increasing spatial curvature at the region, as the volume increases for a fixed boundary area.

\begin{figure}[hbtp]
\begin{centering}
    \includegraphics[width=7.25cm]{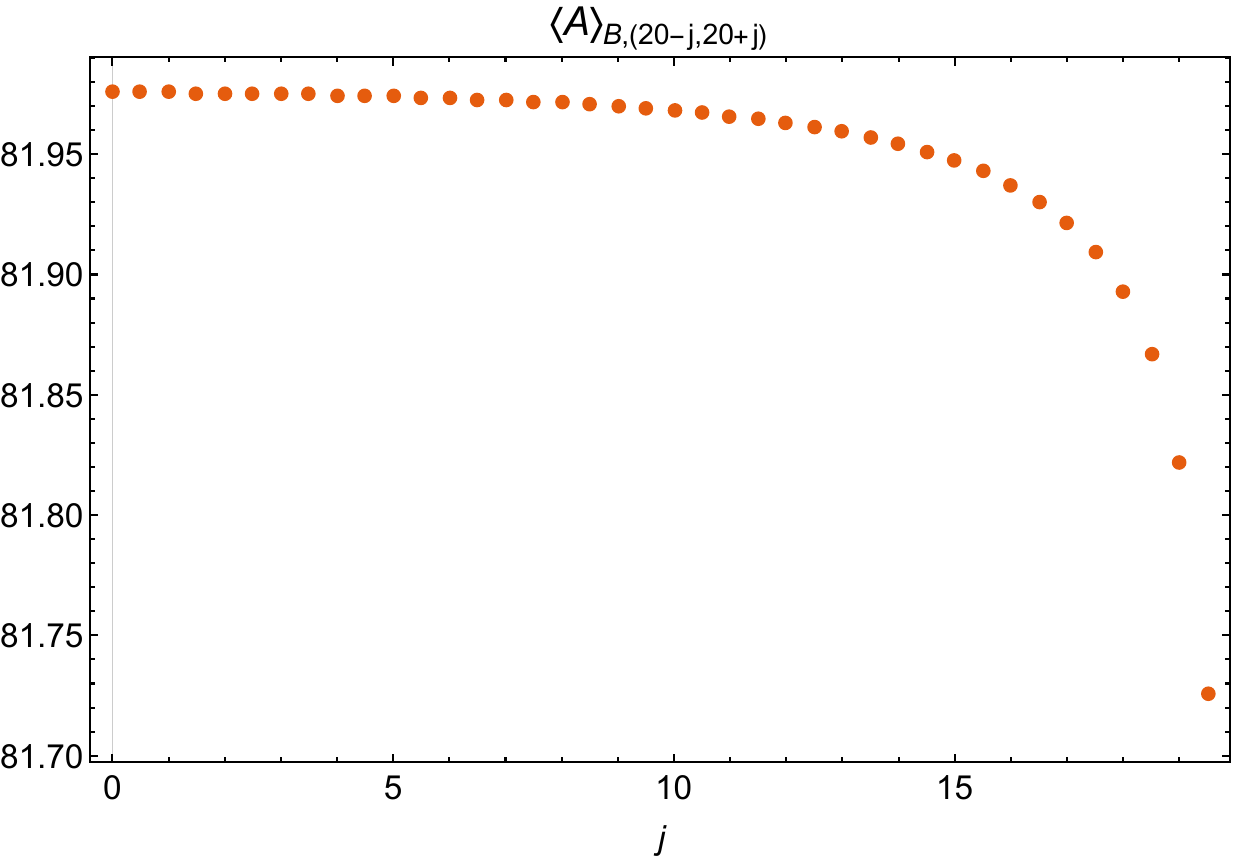}
    \caption{Expectation value $\langle A_S \rangle_{\mathcal{B}, j, J}$ of the total area operator $A_S$ for CBN states $\ket{\Gamma_{2,4}, \mathcal{B},\{j_\ell\}}$, in units of $\mathfrak{a}_{0}$, with $j_1=j_2 = 20 - j, j_3=j_4 = 20 + j$ and $J=80$. Note that $\langle A_S \rangle_{\mathcal{B}, j, J}$ is maximum when $j = J/4$.}
    \label{fig:Asurface}
    \end{centering}
\end{figure}

To complete the description of the geometric observables on a dipole graph, we conclude by studying the expectation value of the surface area $A_S$~(\ref{eq:invariant-local-observable-dipole}) for CBN states on a dipole graph:
\be
\langle A_S \rangle_{\mathcal{B},j_\ell} =  \mathfrak{a}_{0} \sum_{i} \sqrt{j_{i}(j_i+1)}\,, \quad A_S \equiv L \, A \, .
\ee
In Fig.~\ref{fig:Asurface}, we present numerical results for $\langle A_S \rangle_{\mathcal{B},j_\ell}$, for the case of $j_1=j_2 = J/4 - j, j_3=j_4 = J/4 + j$ constrained to a fixed total spin of $J$.

\section{Summary and discussion}
\label{sec:5}

We determined the effective geometry of cosmological Bell-network states on the dipole graph $\Gamma_{2,4}$ through the calculation of expected values and dispersions of the basic geometric observables on this graph, including areas of faces, dihedral angles and the volume at a node. The dipole graph is the simplest graph exhibiting the discrete analogues of homogeneity and isotropy: its nodes, which describe local regions, are indistinguishable from each other, just as its links, which represent discretized directions starting from a node, are indistinguishable among themselves. The dipole graph thus provides a simple combinatorial framework for the representation of the spatial geometry of cosmological spaces in LQG. 

In LQG on a fixed graph, the requirement of diffeomorphism invariance reduces to the residual symmetry condition of automorphism invariance. The requirement that states and observables of the geometry be automorphism-invariant imposes significant restrictions on the space of states and on the algebra of observables, when highly symmetric graphs are considered. We worked exclusively with automorphism-invariant objects to describe the effective geometry. Accordingly, we restricted to invariant states on $\Gamma_{2,L}$, which form the subspace $\mathcal{K}_{\Gamma_{2,L}} \subset \mathcal{H}_{\Gamma_{2,L}}$ characterized by the condition:
\be
U_{\mathrm{A}} \, \ket{\Psi}  = \ket{\Psi}\,, \quad \forall \, \mathrm{A} \in \mathrm{Aut}(\Gamma_{2,L}) \,.
\ee
As shown in~\cite{BY-23}, homogeneous Bell-network states are automorphism-invariant on any graph. As such, they inherit the symmetries of the graph, and on the dipole graph describe discrete homogeneous and isotropic geometries. We denoted such states as cosmological Bell-network (CBN) states $\ket{\Gamma_{2,L},\mathcal{B},\lambda}$. In general, they involve a superposition of states with arbitrary spins. Projecting to states with definite spins $j_\ell$, we obtained CBN states with definite spins $\ket{\Gamma_{2,L}, \mathcal{B},\{j_\ell\}}$. Although the spins $j_\ell$ are not necessarily equal, the CBN state $\ket{\Gamma_{2,L}, \mathcal{B},\{j_\ell\}}$ is isotropic nonetheless, as it includes a superposition of all permutations of such spins. That is, the fluctuations of the geometry include nonisotropic configurations, characterized by the spins $j_\ell$.

We employed autormorphism-invariant operators to describe the geometric observables of interest. Automorphism-invariant observables $\mathcal{O}_{\Sigma, \mathrm{inv}}$ associated with a subgraph $\Sigma$ are obtained by group averaging local operators $\mathcal{O}_{\Sigma}$ that act non-trivially only on a subgraph $\Sigma$:
\be
\mathcal{O}_{\Sigma, \mathrm{inv}} = \frac{1}{2! L!} \sum_{\mathrm{A} \in \mathrm{Aut}(\Gamma_{2,L})} \!\! U_{\mathrm{A}} \, \mathcal{O}_{\Sigma} \, U_{\mathrm{A}}^{-1}\,.
\ee
It turns out that, in any invariant state $\ket{\Psi_{\Gamma_{2,L}}} \in \mathcal{K}_{\Gamma_{2,L}}$, their expectation values are the same as those for the local operators:
\be
\bra{\Psi_{\Gamma_{2,L}}} \, \mathcal{O}_{\Sigma, \mathrm{inv}} \, \ket{\Psi_{\Gamma_{2,L}}} = \bra{\Psi_{\Gamma_{2,L}}} \, \mathcal{O}_\Sigma \,\ket{\Psi_{\Gamma_{2,L}}} \,.
\ee
On $\Gamma_{2,L}$, an elementary set of such invariant observables includes the volume, area, and angle operators $(V, A, \cos \Theta)$, which correspond to invariant one-node, one-link, and one-wedge observables, respectively. 

The average values of these observables define an effective geometry associated with a given state. By numerically computing the effective geometry for several spins $\{j_\ell\}$, as well as the dispersion around it, we have shown that Bell-network states $\ket{\Gamma_{2,L}, \mathcal{B}, \{j_\ell\}}$ with a set of definite spins $\{j_\ell\}$ on $\Gamma_{2,4}$ are not peaked on a classical piecewise linear geometry, such as flat tetrahedra. Instead, their average geometry typically corresponds to that of a regular curved (spherical) tetrahedron. For instance, the mean volume and the dihedral angle between two faces in general do not match those of a regular flat tetrahedron. We found that, in general, the spatial geometry of a CBN state describes the gluing of two identical regular spherical tetrahedra, with perfectly correlated fluctuations of the geometry, and a large relative dispersion of the cosine of the dihedral angle operator. Furthermore, analytical and numerical results indicated that a flat solution of the average geometry can uniquely be realized when all spins are equal, i.e., when $j_\ell = j_0, \forall \ell$. In this case, the average of the cosine of the dihedral angle reads:
\be
D_{\mathcal{B}}(j_0) \equiv \bra{\Gamma_{2,L}, \mathcal{B}, j_0} \cos \Theta \, \ket{\Gamma_{2,L}, \mathcal{B}, j_0} = -\frac{1}{3} \,,
\ee
matching the result for a classical flat tetrahedron. We stress that such a non-trivial description of the local effective geometry, characterized by a curved tetrahedron, is realized at a single node, rather than through a refinement of the graph to achieve a curved geometry.

The results presented herein elucidate the following primary aspects. Firstly, they illustrate how automorphism-invariant tools can be employed for the description of the effective geometry in LQG. General aspects of automorphism-invariance on so-called $2$-CH graphs, which display discrete analogues of homogeneity and isotropy, were discussed in \cite{BY-23}. Here, we concretely implement the general techniques introduced in that work in the simple context of a dipole graph. The automorphism-invariant observables are not associated with specific regions of the graph, as a definite link, for instance, but with equivalence classes of isomorphic regions. An invariant link observable, for instance, describes an observation performed on a single link, which is not specified on the graph, however. Secondly, we considered states that are highly entangled and satisfy an area-law for the entanglement entropy, a feature expected from well-behaved semiclassical states \cite{bianchi-myers}. Moreover, the considered CBN states are not sharply peaked on a classical configuration of the discrete geometry, in constrast with coherent states, for which the effective geometry is simply that of the classical geometry the state is peaked on. For CBN states, the effective geometry cannot in general be described as the gluing of classical flat tetrahedra. On the dipole graph, due to homogeneity and isotropy, the effective geometry is still described as the gluing of two regular tetrahedra, but the relations between the geometric observables are those of more general spherical tetrahedra, and this property is related to the more pronounced fuzziness of the state.

There are several promising directions for further research. A compelling direction involves further extending the application of automorphism-invariant tools beyond the analysis of the effective geometry. In particular, one could investigate how to identify a generic local region and define its entropy in an invariant manner, with respect to invariant states and observables in loop quantum gravity. More specifically, the entropy of a single node has been defined in an invariant manner for a cosmological state on a $2$-CH graph~\cite{BY-23}, and this approach could be extended to more general subregions and contexts. Another direction is the analysis of the effective geometry of Bell-network states on more complex graphs. As remarked before, the statistics of local geometric observables for CBN states display large dispersions. As such, one should not expect that observables associated with regions of a small graph, as a single link or node on a dipole graph with four links, as analyzed, will describe a semiclassical geometry with small fluctuations, but that might still be true for averaged properties on graphs with a large number of individual degrees of freedom. If CBN states can be explored as viable semiclassical states, they must describe a semiclassical geometry only in this sense, and it would be interesting to analyze their effective geometry on larger graphs. In addition, focusing on a class of states that are both highly entangled and invariant with respect to the discrete version of diffeomorphisms might yield a new path to investigate the connection between entanglement and gauge symmetry in quantum gravity~\cite{raamsdonk, ted1, ted2, chirco, cao-carroll, livine-ent, saravani}.


\begin{acknowledgments}
N.Y. acknowledges financial support from the Conselho Nacional de Desenvolvimento Cient\'ifico e Tecnol\'ogico (CNPq) under Grant No.~306744/2018-0. 
\end{acknowledgments}

\appendix

\section{Geometry of non-Euclidean tetrahedra}
\label{sec:appendix}

A spherical tetrahedron is a region in the unit sphere $\mathbb{S}^3$ bounded by four flatly embedded surfaces intersecting only at their boundaries. It has four vertices connected by edges that are great arcs in the sphere, and its faces are parts of great two-spheres. Similarly, a hyperbolic tetrahedron is a region of the hyperbolic space $\mathbb{H}^3$ bounded by four flatly embedded surfaces intersecting only at their boundaries \cite{haggard}. A non-Euclidean tetrahedron can be either a spherical or a hyperbolic tetrahedron. Properties of non-Euclidean tetrahedra are reviewed in \cite{abrosimov}, to which we refer for more details. We are mostly interested in the case of spherical tetrahedra.

A concrete presentation of a spherical tetrahedron can be built as follows \cite{kolpakov}. Let $\mathbb{R}^4$ be the four-dimensional Euclidean space and $\mathbb{S}^3=\{x \in \mathbb{R}^4 : \delta_{ij} x^i x^j =1 \}$ be the unit sphere embedded as the boundary of a $4$-ball of unit radius in $\mathbb{R}^4$. Let $v_1, \dots, v_4$ be four vectors in $\mathbb{R}^4$. The cone over $\{v_i\}$ is defined as
\be
\mathrm{cone}(v_1, \dots, v_4)=\left\{ \sum_{i=1}^4 \lambda_i v_i : \lambda_i \geq 0, i=1, \dots, 4 \right\} \, .
\ee
A spherical tetrahedron is the intersection of a cone over four linearly independent vectors $v_1, \dots, v_4$ and the unit sphere in $\mathbb{R}^4$. 

\begin{figure}
\begin{centering}
\begin{tikzpicture}[scale=0.85]
\filldraw [black] (-0.1,0.15) circle [radius=2.5pt];
\filldraw [black] (-2,2) circle [radius=2.5pt];
\filldraw [black] (0.93,4.9) circle [radius=2.5pt];
\filldraw [black] (2.53,1.7) circle [radius=2.5pt];
\draw (-2,2) arc [start angle=-150, end angle=-120, radius=5];
\draw (-2,2) arc [start angle=152, end angle=118, radius=7];
\draw (-2,2) arc [start angle=100, end angle=72.5, radius=9.5];
\draw (-0.1,0.15) arc [start angle=-78, end angle=-40, radius=4.8];
\draw (-0.1,0.15) arc [start angle=177, end angle=158, radius=14.8];
\draw (-0.1,0.15) arc [start angle=177, end angle=158, radius=14.8];
\draw (0.93,4.91) arc [start angle=40, end angle=13, radius=7.63];
\draw (3.2,1.65) node {1};
\draw (-0.12,-0.5) node {2};
\draw (-2.7,2.05) node {3};
\draw (1,5.5) node {4};
\draw (-1.2,3.8) node {\footnotesize A};
\draw (-0.1,3) node {\footnotesize B};
\draw (2.2,3.8) node {\footnotesize C};
\draw (1.75,.45) node {\footnotesize D};
\draw (1.5,2.3) node {\footnotesize E};
\draw (-1.65,.8) node {\footnotesize F};
\end{tikzpicture}
\end{centering}
\caption{Vertices and dihedral angles in a non-Euclidean tetrahedron.}
\label{fig:curved-tetrahedron}
\end{figure}
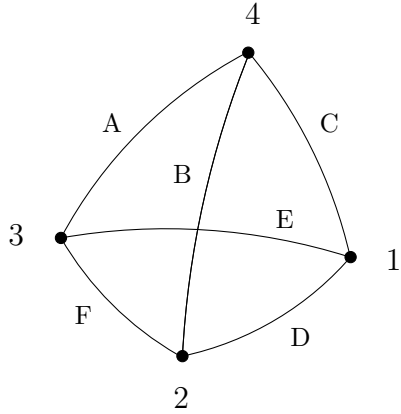

A non-Euclidean tetrahedron is uniquely determined, up to isometries, by its dihedral angles. In constrast, a flat tetrahedron is specified by its dihedral angles only up to similarity. Let the vertices $1,\dots,4$ and the (interior) dihedral angles $\mathrm{A},\mathrm{B},\mathrm{C},\mathrm{D},\mathrm{E},\mathrm{F}$ be related as represented in Fig.~\ref{fig:curved-tetrahedron}. The Gram matrix of the tetrahedron is defined as
\be
G= \begin{pmatrix}
1		& -\cos \mathrm{A} & -\cos \mathrm{B} & -\cos \mathrm{F} \\
-\cos \mathrm{A} & 1	 & - \cos \mathrm{C} & -\cos \mathrm{E} \\
-\cos \mathrm{B} & -\cos \mathrm{C} & 1 & -\cos \mathrm{D} \\
-\cos \mathrm{F} & -\cos \mathrm{E} & -\cos \mathrm{D} & 1
\end{pmatrix}.
\label{eq:Grammatrix}
\ee
We denote the coefficients of the Gram matrix by $G_{ij}$, and the coefficients of its cofactor matrix $\Cof(G)$ by $c_{ij}$. An information that can be easily extracted from the Gram matrix is whether a tetrahedron is flat, spherical or hyperbolic:
\be
\det G: \quad
\begin{cases}
<0 \, , & \text{hyperbolic} \, , \\
=0 \, ,& \text{Euclidean} \, , \\
>0 \, , & \text{spherical} \, .
\end{cases}
\label{sign-detG}
\ee

The edge lengths are related in a simple manner to the coefficients of $\Cof(G)$. Let $\ell_{ij}$ be the length of the edge connecting the vertices $i$ and $j$. For a spherical tetrahedron, $c_{ii}>0$, and
\be
\qquad \cos \ell_{ij} = \frac{c_{ij}}{\sqrt{c_{ii} c_{jj}}} \quad \text{(spherical tetrahedron)} \, .
\ee
For a hyperbolic tetrahedron, $c_{ij}>0$, and
\be
\qquad \cosh \ell_{ij} = \frac{c_{ij}}{\sqrt{c_{ii} c_{jj}}} \quad \text{(spherical tetrahedron)} \, .
\ee

The volume of a non-Euclidean tetrahedron satisfies the differential Schl\" affi formula \cite{abrosimov}. The variation of the volume $V_{\mathrm{sph}}$ is related to the variation of the dihedral angles through the relation:
\be
2 K dV_{\mathrm{sph}} = \sum_{i<j} \ell_{ij} \, dG_{ij} \, ,
\ee
where $K$ is the sectional curvature of the space, i.e., $K=+1$ for a spherical tetrahedron on $\mathbb{S}^3$ of unit radius ($K=-1$ for a hyperbolic tetrahedron on $\mathbb{H}^3$). An integral representation of the volume that follows from the Sch\"affi formula is known as the Sforza's formula \cite{abrosimov}. An explicit formula for the volume written in terms of dilogarithm functions was obtained in \cite{murakami}. The special case of symmetric tetrahedra satisfying $\mathrm{A}=\mathrm{D}$, $\mathrm{B}=\mathrm{E}$ and $\mathrm{C}=\mathrm{F}$ was studied in \cite{derevnin}.

\begin{figure}
    \centering
    \includegraphics[width=7cm]{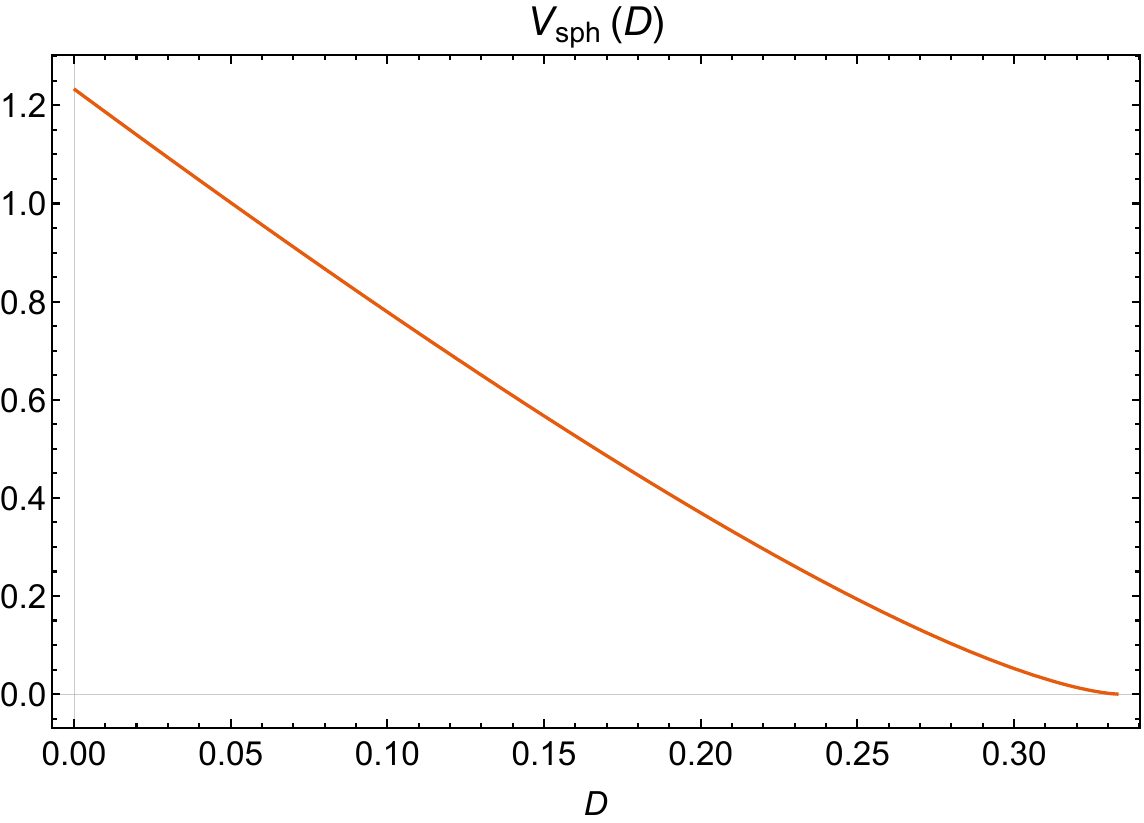}
    \caption{The volume $V_{\mathrm{sph}}(D)$~\eqref{eq:Vclassical} of a regular spherical tetrahedron in terms of the cosine of the (interior) dihedral angle, $D \equiv \cos \mathrm{A}$.}
    \label{fig:Vclassical}
\end{figure}

We are interested in the special case of regular spherical tetrahedra. Let us then set all dihedral angles to be equal: $G_{ij}=- \cos\alpha$, for $i \neq j$. In this case,
\be
\det G = 1- 6 \cos^2 \alpha - 8 \cos^3 \alpha - 3 \cos^4 \alpha \, ,
\ee
and the cofactor matrix has coefficients
\be
c_{ij} = \begin{cases}
	1 - 3 \cos^2 \alpha - 2 \cos^3 \alpha \, , & i=j \, , \\
	\cos \alpha + 2 \cos^2 \alpha + \cos^3 \alpha \, , & i \neq j \, .
\end{cases}
\ee
In addition, for a regular spherical tetrahedron on $\mathbb{S}^3$, the edge lengths $\ell_{ij}=\ell$ can be expressed in terms of the dihedral angle as
\be
\ell = -\frac{i}{2} \log \left[ \frac{\cos \alpha + i \sqrt{3(\cos \alpha -1)(\cos \alpha - 1/3)}}{\cos \alpha - i \sqrt{3(\cos \alpha -1)(\cos \alpha - 1/3)}} \right] \, .
\ee

The faces of a spherical tetrahedron are equilateral spherical triangles on $\mathbb{S}^2$. The internal angles $\beta$ are related to the edge lengths by
\be
\cos \beta = \frac{\cos \ell}{1+\cos \ell} \, ,
\ee
and the area of a face is
\be
A_f = 3\beta-\pi \, .
\label{eq:area-face}
\ee

\begin{figure}[hbtp]
    \centering
    \includegraphics[width=7cm]{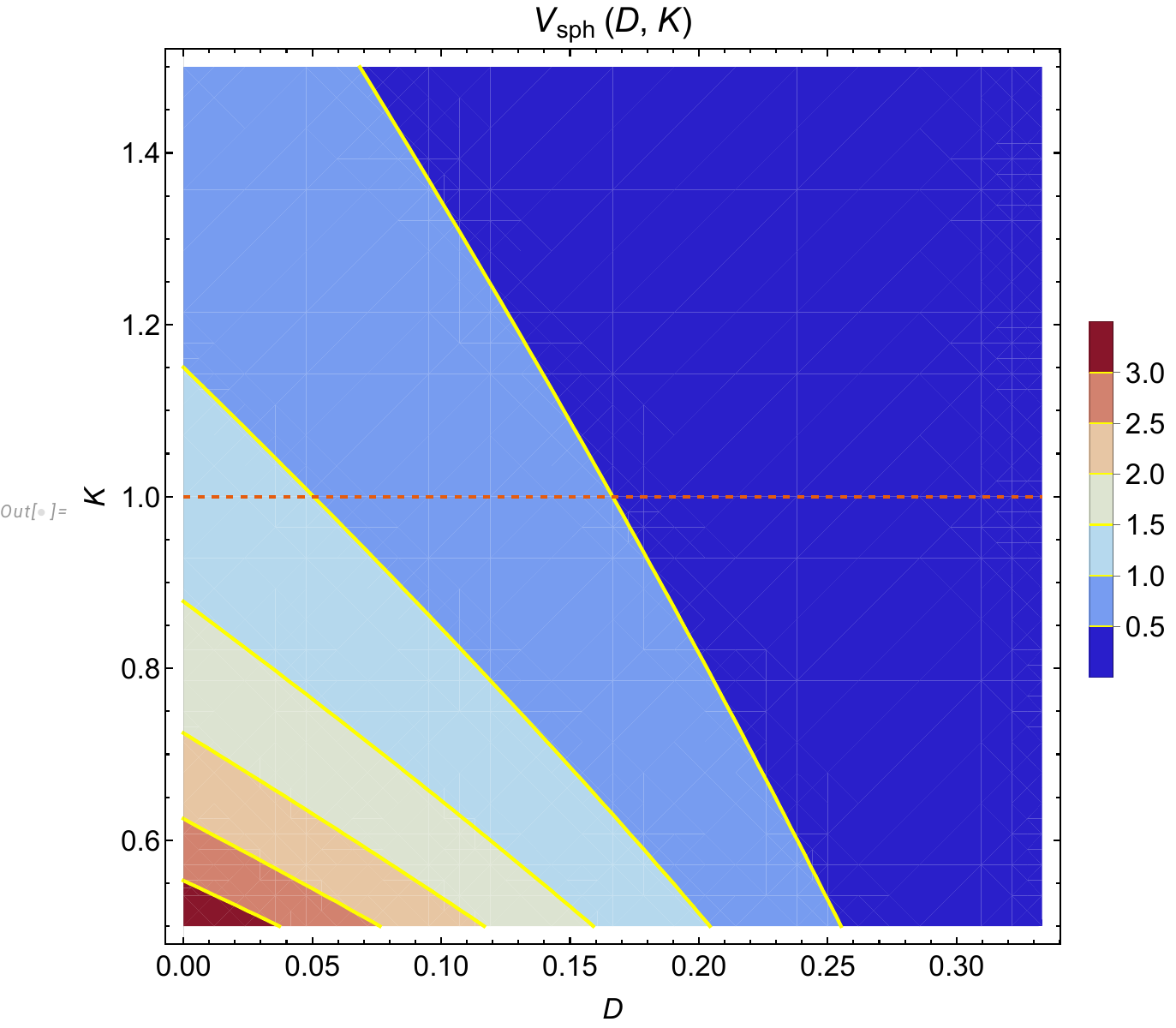}
    \caption{The volume $V_{\mathrm{sph}}(D,K) \equiv K^{-\frac{3}{2}} \, V_{\mathrm{sph}}(D)$ of a regular spherical tetrahedron on a 3-sphere in terms of the sectional curvature $K$ and the cosine of the (interior) dihedral angle $D$.}
    \label{fig:VclassicalK}
\end{figure}

The volume of a regular spherical tetrahedron with $K=1$ is given in a integral form by \cite{derevnin}:
\be
V_{\mathrm{sph}}(D) = -\int_v^{\infty} \left( 3 \arcsinh \frac{D}{\sqrt{\nu^2-1}} - \arcsinh \frac{1}{\sqrt{\nu^2-1}} \right) \frac{1}{\nu}d\nu \, ,
\label{eq:Vclassical}
\ee
where
\be
v=\frac{1- 3 D^2 - 2 D^3}{\sqrt{(1+D)^3 (1-3 D)}} \,, \quad D \equiv \cos \mathrm{A}
\ee

It is noteworthy that $V_{\mathrm{sph}} \to 0$ as $D \to \frac{1}{3}$, signifying the transition of the tetrahedron to a regular flat tetrahedron. In Fig.~\ref{fig:Vclassical}, we report numerical exploration of the volume of a regular spherical tetrahedron as a function of $D$.

If a spherical tetrahedron is embedded on a sphere $\mathbb{S}_a^3$ of non-unit radius $a$, the lengths, areas and volume must be rescaled accordingly through multiplication by factors of $a$, $a^2$ and $a^3$ or analogously by $K^{-1/2}$, $K^{-1}$ and $K^{-3/2}$, respectively (See Fig.~\ref{fig:VclassicalK}). 



\end{document}